\documentclass[10pt,journal, twocolumn]{IEEEtran}
\usepackage{cite}
\usepackage{amsmath,amssymb,amsfonts,bm}
\usepackage{dsfont}
\usepackage{algorithmic}
\usepackage{graphicx}
\usepackage{textcomp}
\usepackage{booktabs}
\usepackage{xcolor}
\usepackage{color}
\usepackage{tabularx}
\usepackage{algorithm}
\usepackage{algorithmic}
\usepackage[T1]{fontenc}
\def\BibTeX{{\rm B\kern-.05em{\sc i\kern-.025em b}\kern-.08em
    T\kern-.1667em\lower.7ex\hbox{E}\kern-.125emX}}

\graphicspath{{./fig/}}
\newcommand{\qa}{{\bf a}}
\newcommand{\qb}{{\bf b}}
\newcommand{\qc}{{\bf c}}

\newcommand{\qe}{{\bf e}}

\newcommand{\qg}{{\bf g}}
\newcommand{\qh}{{\bf h}}

\newcommand{\qm}{{\bf m}}

\newcommand{\qp}{{\bf p}}
\newcommand{\qq}{{\bf q}}

\newcommand{\qs}{{\bf s}}

\newcommand{\qw}{{\bf w}}

\newcommand{\qA}{{\bf A}}

\newcommand{\qE}{{\bf E}}

\newcommand{\qH}{{\bf H}}
\newcommand{\qI}{{\bf I}}

\newcommand{\qW}{{\bf W}}

\newcommand{\be}{\begin{equation}} \newcommand{\ee}{\end{equation}}
\newcommand{\bea}{\begin{eqnarray}} \newcommand{\eea}{\end{eqnarray}}

\begin{document}

\title{A Data and Model-Driven Deep Learning Approach to Robust Downlink Beamforming Optimization}
\author{Kai Liang,
        Gan Zheng,~\IEEEmembership{Fellow,~IEEE,} Zan Li,~\IEEEmembership{Senior Member,~ IEEE},\\ Kai-Kit Wong,~\IEEEmembership{Fellow,~IEEE}, and Chan-Byoung Chae,~\IEEEmembership{Fellow,~IEEE}
              \thanks{K. Liang, and Z. Li  are with the State Key Laboratory of Integrated Service Networks, the School of Telecommunications Engineering, Xidian University, Xi'an, 710071, China  (email: { kliang@xidian.edu.cn, zanli@xidian. edu.cn}).}
     \thanks{G. Zheng is  with the School of Engineering, University of Warwick, Coventry, CV4 7AL, UK
 (email: gan.zheng@warwick.ac.uk).}
  \thanks{K.-K. Wong is with the Department of Electronic and Electrical Engineering, University College London, London, WC1E 6BT, UK (email: kai-kit.wong@ucl.ac.uk). He is also affiliated with Yonsei Frontier Lab., Yonseu University, Seoul, 03722, Korea.}
  \thanks{C.-B. Chae is with School of Integrated Technology, Yonsei University, Seoul, 03722, Korea (email: cbchae@yonsei.ac.kr).}
}
\maketitle
\begin{abstract}
This paper investigates the optimization of the probabilistically robust transmit beamforming problem with channel uncertainties in the multiuser multiple-input single-output (MISO) downlink transmission. This problem poses significant analytical and computational challenges. Currently, the state-of-the-art optimization method relies on convex restrictions as tractable approximations to ensure robustness against Gaussian channel uncertainties. However, this method not only exhibits high computational complexity and suffers from the rank relaxation issue but also yields conservative solutions. In this paper, we propose an unsupervised deep learning-based approach that incorporates the sampling of channel uncertainties in the training process to optimize the probabilistic system performance. We introduce a model-driven learning approach that defines a new beamforming structure with trainable parameters to account for channel uncertainties. Additionally, we employ a graph neural network to efficiently infer the key beamforming parameters. We successfully apply this approach to the minimum rate quantile maximization problem subject to outage and total power constraints. Furthermore, we propose a bisection search method to address the more challenging power minimization problem with probabilistic rate constraints by leveraging the aforementioned approach. Numerical results confirm that our approach achieves non-conservative robust performance, higher data rates, greater power efficiency, and faster execution compared to state-of-the-art optimization methods.
 \end{abstract}


\section{Introduction}\label{Sec: Intro}
\IEEEPARstart{B}{eamforming} techniques in multi-user and multi-antenna systems can effectively improve the throughput and spectral efficiency of the wireless communication system, and have received long-term attention from both academia and industry \cite{BF0,SPM1,BF1,BF2}.
Existing studies utilize beamforming algorithms with known channel state information (CSI) at the base station (BS) and perform interference cancellation and resource allocation among multiple users.
However, perfect CSI in practical scenarios is difficult to obtain, due to the presence of channel estimation errors, the quantization errors from the user side feedback, and the time-varying nature of wireless channels.

Since the imperfect CSI can worsen the performance of multi-user multi-antenna system, this has motivated researchers to carry out the robust design of beamforming,
i.e., the optimal beamforming design based on predefined CSI error model to mitigate the negative impact of imperfect CSI.
Existing works can usually be divided into three main categories,
namely, the worst-case robust approach \cite{WorstCase,WorstCase2,WorstCase3}, the average robust approach \cite{AverageBobust,AverageBobust2}, and the outage-constrained approach \cite{outage1,outage2,tsp-solution}.
The worst-case robust approach aims to guarantee the worst-case performance over all the potential channel errors which are assumed to be bounded within some closed sets.
Since the worst-case scenario is a rarely happening event in a multi-antenna system, the solutions obtained by this approach are usually very conservative and requires high resource consumptions.
The average robust approach aims to optimizing the average performance, but fails to guarantee the outage requirement,
which is vital for  delay- or reliability- sensitive applications in the next generation communication system.
Thus, in this paper, we focus on the outage-constrained approach,
because it is able to guarantee the probabilistic  quality of service (QoS)  requirements of   users.
This approach focuses on the QoS outage probability constraint based on a probabilistic CSI error model (such as the Gaussian model).

Solving the outage constrained beamforming optimization problem appears very hard,
and the key challenge is how to deal with the outage probability which in general does not have a  closed-form expression.
Traditionally, the outage constrained beamforming problem can be solved by convex optimization after approximating the rate outage constraints.
In \cite{tsp-solution},  the authors provided two approximation methods (i.e., Bernstein-type inequality (BTI) method and large deviation inequality (LDI) method) together with semideﬁnite relaxation \cite{SDP1,SDP2} under Gaussian CSI uncertainties to solve the outage constrained robust beamforming problem and they are still known to achieve  the best solution so far.
These methods have been widely used in the studies of robust beamforming  design with CSI uncertainties, such as intelligent reflecting surface \cite{irs}, simultaneous wireless information and power transfer (SWIPT)  \cite{swipt}, heterogeneous networks \cite{HetNet}, satellite-based Internet of Things \cite{Satellite}, and radar-communication systems \cite{Rada-communication}.
Nonetheless, these methods have a high computational complexity and algorithmic latency caused by, for example, iterative algorithms and complex matrix computations.
In addition, since these methods attempt to approximate the outage constraint by finding the upper bound of outage probability, they usually yield sub-optimal and conservative solutions as well.
Finally, for multi-user multi-input single-output (MU-MISO)  systems, these methods may obtain high-rank solutions due to  semidefinite relaxation, which cannot satisfy the rank-1 constraint and cannot be decomposed into beamforming vectors.
Although Gaussian randomization procedure \cite{SDP1} may deal with this problem, it further weakens the optimality of the solution, increases the computational and delay overhead, and even obtains non-feasible solutions for the outage constraint.

In recent years, deep learning-based beamforming approaches have been increasingly investigated.
It uses deep neural networks (DNNs) to approximate the mapping of inputs and optimal solutions of traditional optimization algorithms, thus bypassing the tedious iterative solution process and handling non-convex optimization problems with ease.
Compared to purely data-driven methods, the data- and model-driven learning method can effectively improve the training efficiency and performance of neural networks, and has a wide range of applications in the field of wireless communications, such as transceiver design in over-the-air federated learning \cite{FedLearning}, and computation ofﬂoading in mobile-edge computing networks \cite{Offloading}.
As for beamforming optimization, we proposed a model-driven learning approach for MU-MISO  systems with the aid of a convolutional neural network (CNN) in \cite{dl_bf},
where the CNN's output dimension is reduced by adopting an optimal beamforming structure \cite{bf-structure}.
We further used this method for the case of the absence of uplink and downlink channel reciprocity as in \cite{Model_Driven}.
This approach is then widely extended for applications in DNN-based optimal beamforming designs such as universal power MISO beamforming \cite{Universal-power}, scalable beamforming optimization though graph neural network \cite{GNN},  and multi-user mutli-input mutli-output precoding \cite{MU-MIMO}.
In addition to optimal beamforming structures, other types of mathematical model-driven methods are also used in the design of beamforming based on perfect CSI.   In \cite{ModelNew1}, a deep unfolding network was proposed, which unfolds the weighted minimum mean-square error (WMMSE) algorithm \cite{WMMSE} for solving problem sum rate maximization problems and learns beamforming matrices through neural networks. The method was further extended to the design of hybrid beamforming in millimeter-wave systems \cite{ModelNew2}. In \cite{ModelNew3}, a model GNN was proposed by using first-order Taylor's expansion to approximate inverse matrices and achieve better performance.
However, these methods are designed with perfect CSI at the transmitters, and cannot directly applied for robust beamforming optimization.

Meanwhile, many works have emerged for the deep learning based robust beamforming optimization.
The authors in \cite{StatisticRobust} proposed a DNN-based robust beamforming approach to solve average sum rate maximization problem in MU-MISO system under Gaussian channel uncertainties.
In \cite{Multicast-unicast}, a graph neural network (GNN) based robust beamforming method was proposed with Gaussian channel uncertainties.
Nevertheless, these works did not consider the outage probability performance.
In \cite{icc}, we proposed a data augmentation based DNN approach to obtain robust beamforming for the outage constrained power minimization problem, which achieves better power efficiency and the execution delay.
However, this method directly learns beamforming vectors without dimensional reduction through the optimal structure, making it hard to train the neural network.
A similar method was presented for MU-MISO to address the outage-constrained rate quantile maximization problem in \cite{power-only-solution}, where beamforming vectors are recovered by the learned power features and the regularized zero-forcing (RZF) \cite{RZF,CBF} beamforming structure.
This scheme reduces the dimensionality of the output and makes the neural network easy to train, but sacrifices the transmission performance due to the inferior performance of RZF beamforming.

In this paper, we explore the application of data and model-driven deep learning as a novel approach to address two robust beamforming optimization problems in MU-MISO systems. Our proposed method exhibits improved rate and power performance in the presence of channel uncertainty. Additionally, we achieve lower output dimensions to enhance neural network training efficiency and employ a more flexible neural network structure to accommodate universal transmit power. To the best of our knowledge, our work is the first to surpass the performance and computational efficiency of the traditional solution presented in \cite{tsp-solution}.
The contributions are summarized in the following:

\begin{itemize}
	\item We propose a data and model-driven deep learning approach to solve the outage-constrained minimum rate quantile maximization problem, while adhering to the rate outage probability constraint and the transmit power constraint, considering Gaussian channel uncertainties.
	We first propose a modified optimal beamforming structure that introduces new trainable interference features arising from channel estimation errors. These features, along with the power features, enhance system performance while simultaneously reducing the neural network's output dimension.
	Furthermore, we introduce a specialized GNN, as described in \cite{GNN}, to train interference and power features for improved rate performance, scalability and generalizability of the neural network.
	Unlike the  message passing GNN \cite{MPGNN} that can only handle homogeneous graphs, this GNN realizes the bipartite message-passing (BMP) inference, which makes it capable of handling heterogeneous vertex types  using different neural networks.
	After implementing beamforming recovery using the modified optimal structure, we estimate the rate quantile through data augmentation and Monte Carlo (MC) sampling.  MC sampling is unnecessary for the inference of robust beamforming, ensuring that it does not increase the execution time or complexity.
	\item We further propose a bisection-based deep learning method to address the robust beamforming design for the more challenging transmit power minimization problem  with the rate outage probability constraint.
	Existing neural network methods, such as those discussed in \cite{icc}, pose challenges in training as they require simultaneous updating of the neural network parameters and Lagrange multipliers. 
	To overcome this limitation, we modify the neural network framework designed for rate quantile maximization. We incorporate transmit power as an additional input, enabling the neural network to feature universal transmit power. The trained universal neural network, in conjunction with the bisection method, allows us to infer the minimum transmit power that satisfies the outage probability constraints. This leverages the observation that the transmission rate increases with higher transmit power.
	\item Numerical results show that our proposed robust beamforming method outperforms state-of-the-art  works, including the classical optimization method \cite{tsp-solution},  the RZF-model based learning  method \cite{power-only-solution} and the DNN-based learning method without assistance of model information,  in terms of achievable robust rate, execution time, power consumption  and feasibility.
\end{itemize}

The rest of this paper is organized as follows.
System model and problem formulation are described in Section \ref{Sec:SystemModule},
after which the model-driven learning approach for the rate quantile maximization problem is provided in Section \ref{Sec: RateQuantile}.
Then, the bisection based deep learning approach for the power minimization problem is given in Section \ref{Sec:PowerMin}.
Numerical results and analysis are provided in Section \ref{Sec:Results}, followed by which we conclude this paper in Section \ref{Sec:Conclusion}.

\textbf{Notation}:
Bold uppercase and lowercase letters (e.g., $\qA$ and $\qa$) represent matrices and column vectors, respectively.
$\qI$ denotes the identity matrix.
The superscripts  $(\cdot)^T$ and $(\cdot)^\dag$ stand for transpose and Hermite conjugate transpose of a matrix or vector, respectively.
We use $\mathbb{R}^{m\times n}$ and $\mathbb{C}^{m\times n}$ to denote the real and complex space of m by n dimensions, respectively.
$Re(\cdot)$ and $Im(\cdot)$ represent the real and imaginary parts of a complex variable, respectively.
$|\cdot|$ and $\Vert \cdot \Vert$ refer to the absolute value operator and the Euclidean norm operator, respectively.
$\lfloor \cdot \rfloor$ and $\lceil \cdot \rceil$  denote the floor and ceil functions (i.e. the downward  and upward rounding operations), respectively.
$dim(\qa)$ represents the dimension of a vector $\qa$.

\section{System Model and Problem Formulation}\label{Sec:SystemModule}
We consider a  MU-MISO   downlink system  in which a BS with  $N$-antennas serves $K$ single-antenna users.
Let $\mathcal{N}=\{1,\cdots,N\}$ and $\mathcal{K}=\{1,\cdots,K\}$ denote the antenna set and the user set, respectively.
Suppose $x_k$ is the transmit signal to the user $k$ with unit power  and the BS transmits with a total power $P$.
 The transmitted data symbol $x_k$   is mapped onto the antenna array elements by multiplying the beamforming vector $\mathbf{w}_k \in \mathbb{C}^{N\times 1}$. The received signal at the user $k$ can be expressed as
\begin{align}\label{sys1}
y_k= \mathbf{h}_{k}^\dag \mathbf{w}_k x_k  + \sum_{j\neq k}\mathbf{h}_k^\dag \mathbf{w}_jx_j +n_k,
\end{align}
where $\mathbf{h}_{k} = [h_{1k},\cdots,h_{nk},\cdots,h_{Nk}]^T \in \mathbb{C}^{N\times 1}$, and $h_{nk}$ is the channel between the antenna $n$ at the BS and the user $k$.
Then, the overall channel matrix can be denoted as $\qH = [\qh_1,\cdots,\qh_K]$.
$n_k$ denotes the additive white Gaussian noise (AWGN) component with zero mean and variance $\sigma^2_k$. Therefore, the signal-to-interference-plus-noise ratio (SINR) that measures quality of the data detection   at the $k$-th user is given by
\begin{equation}
\Gamma_k= \frac{  |\mathbf{h}_k^\dag \mathbf{w}_k|^2}{\sum_{j\neq k} |\mathbf{h}_{k}^\dag\mathbf{w}_j|^2+\sigma^2_k},
\label{eq:Gamma}
\end{equation}
and the corresponding achievable rate is formulated as
\begin{equation}\label{eq: rate}
	R_k =B \log_2 (1+\Gamma_k),
\end{equation}
 where $B$ denotes the bandwidth.

Considering that it is hard to obtain perfect CSI at the BS in practice,  the channel error model is given as follows:
\begin{equation}\label{eq: realChannel}
	\mathbf{h}_k = \mathbf{\tilde{h}}_k + \mathbf{e}_k,
\end{equation}
where $\mathbf{\tilde{h}}_k$ is the estimated CSI known at the BS,
and $\mathbf{e}_k = [e_{1k},\cdots,e_{nk},\cdots,e_{Nk}]^T \in \mathbb{C}^{N\times 1}$ represents the channel error.
The overall estimated channel matrix and channel error matrix are denoted as $\tilde{\qH} = [\tilde\qh_1,\cdots,\tilde\qh_K]$ and ${\qE} = [\qe_1,\cdots,\qe_K]$, respectively.
The commonly used Gaussian channel error is adopted to model $\mathbf{e}_k$,
where each element of  $\mathbf{e}_k$ independently identically distributed  follows $\mathcal{CN}(0,\sigma_e^2)$.

 We first focus on the  minimum rate quantile maximization problem,
 i.e., maximizing the corresponding quantile while satisfying the minimum rate outage probability constraint and the BS transmit power constraint.
 Note that the minimum rate (denoted as $R_{min} = \min_{k \in \mathcal{K} } R_k$) is the minimum achievable rate among all users.
 This problem is given as follows:
\begin{eqnarray}
	\textbf{P1:} && \max_{\mathbf{w}_k,\forall \mathcal{K}} r\\
	\mbox{s.t.} && \text{Prob}\big (R_{min}\leq r \big) \leq \rho \label{eq: OutageConstraint}\\
	&& R_{min} \leq R_k, \forall k\in \mathcal{K}\\
	 && \sum_{k=1}^{K} \Vert \mathbf{w}_k \Vert^2 \leq P \label{eq:PowerConstaint},
\end{eqnarray}
where \eqref{eq: OutageConstraint} refers to the outage constraint, and $\rho$ is the target outage probability which often has a low value (e.g., 5\%).
\eqref{eq:PowerConstaint} denotes the power constraint at the BS and $P$ is the maximum transmit power of the BS.

Another conventional but more challenging problem considered in this paper is the power minimization problem based on the minimum rate outage probability constraint,
which is given by 
\begin{eqnarray}
	\textbf{P2:} && \min_{\mathbf{w}_k,\forall \mathcal{K}} \sum_{k=1}^{K} \Vert \mathbf{w}_k \Vert^2 \\
	\mbox{s.t.} &&  \text{Prob}\big (R_{min}\leq r \big) \leq \rho \notag \\
	&& R_{min} \leq R_k, \forall k\in \mathcal{K} .\notag
\end{eqnarray}

Obviously, the optimization problems \textbf{P1} and \textbf{P2} are both nonconvex and difficult to solve.
The traditional method solves such problems through jointly using convex restriction approach and semi-positive definite optimization  (see \cite{tsp-solution}).
However, it suffers from a long convergence time and high-rank  solutions.
Although the Gaussian randomization procedure \cite{SDP1} claims to be able to solve the high-rank matrix issue which  is not designed for the outage constrained problem, it is shown in our simulations that this method can yield infeasible solutions to the above outage-constrained problems with a higher probability.
Therefore, we will present the proposed data and model-driven deep learning approach to solve \textbf{P1} and \textbf{P2}, achieving faster convergence and better performance.

\section{Rate Quantile Maximization}\label{Sec: RateQuantile}
In this section, we provide a deep leaning approach for the rate quantile maximization problem (i.e., \textbf{P1}) via an unsupervised learning manner.
Suppose there exists a function $\mathbf \Psi_1(\cdot)$  with its input being the estimated channel information and its output being an optimal robust beamforming solution that satisfies the constraints in problem \textbf{P1}. Then, the problem  \textbf{P1} can be rewritten as follows:
\begin{equation}\label{eq: P1map}
	\{\qw_1,\cdots,\qw_K\} = \mathbf{\Psi}_1(\tilde{\qh}_1,\cdots,\tilde{\qh}_K).
\end{equation}
Normally, the closed-form expression for function $\mathbf \Psi_1(\cdot)$ does not exist.
Fortunately, the universal approximation theorem \cite{UniversalApproximation} has verified that any continuous-valued functions can be approximated with very small approximation error through a well-designed DNN.
Thus, the original problem can be turned into a trainable problem.
Assume there is a DNN $\mathbf \Psi_1 (\cdot; \mathbf{\Theta})$ with trainable parameter set $\mathbf{\Theta}$, which has the ability to approximate \eqref{eq: P1map} as follows:
\begin{equation}
	\{\qw_1,\cdots,\qw_K\} = \mathbf{\Psi}_1(\tilde{\qh}_1,\cdots,\tilde{\qh}_K; \mathbf{\Theta}).
\end{equation}
This allows us to learn the parameter set $\mathbf{\Theta}$ to obtain robust solutions for the problem \textbf{P1},
which is more tractable than the original problem.

Different from commonly used data driven learning, we propose a model-based beamforming structure to deal with channel uncertainties, which together with data samples speeds up  training  the neural network.
Then, we employ a unique GNN  combined with  MC sampling to realize a robust beamforming optimization while satisfying the outage probability constraint and the power constraint.

\subsection{Model-based Beamforming Structure}
With the known CSI at the BS, we have previously proposed a hybrid supervised and model-based deep learning method in  \cite{Model_Driven} with the aid of the optimal beamforming structure \cite{bf-structure}, which is given below
\begin{equation}\label{eq:w_perfect}
	\mathbf{w}_k = \sqrt{p_k} \frac{\big(\mathbf{I} + \sum_{j=1}^K \frac{q_j}{\sigma_j^2} \mathbf{h}_j \mathbf{h}_j^\dag \big)^{-1} \mathbf{h}_k}{\bigg \Vert\big(\mathbf{I} + \sum_{j=1}^K \frac{q_j}{\sigma_j^2} \mathbf{h}_j \mathbf{h}_j^\dag \big)^{-1} \mathbf{h}_k \bigg \Vert},
\end{equation}
where $p_k$, $q_k$ are non-negative trainable parameters and satisfy $\sum_{k=1}^{K} p_k = \sum_{k=1}^{K} q_k  = P$.
$\mathbf{p}=[p_1,\cdots,p_K]^T \in\mathbb{R}^{K\times1}$ denotes the downlink power allocation vector, and $\mathbf{q}=[q_1,\cdots,q_K]^T\in \mathbb{R}^{K\times1}$ is an auxiliary variable used to determine the direction of optimal beamforming.

This practice reduces the neural network output dimension and thus improves the training efficiency and accuracy.
Unfortunately, this optimal structure cannot be directly applied to the case of channel uncertainty due to the fact that the neural network needs to reconstruct the beamforming vectors based on the known  CSI and the learned power parameters.
To this end, we modify the optimal beamforming structure in order to make it suitable for the case of imperfect CSI.
Specifically, we introduce a key feature implying the extra interference arising from channel estimation errors into the optimal beamforming structure \eqref{eq:w_perfect}.
The modified robust beamforming structure is given as follows:

\begin{equation}\label{eq:w}
	  \hat\qw_k = \sqrt{p_k} \frac{\bigg((1+s_k)\mathbf{I} + \sum_{j=1}^K \frac{q_j}{\sigma_j^2} \tilde\qh_{j}\tilde\qh_{j}^\dag  \bigg)^{-1} \tilde \qh_k}{\bigg \Vert\bigg((1+s_k)\mathbf{I} + \sum_{j=1}^K \frac{q_j}{\sigma_j^2} \tilde\qh_{j}\tilde\qh_{j}^\dag  \bigg)^{-1} \tilde \qh_k \bigg \Vert},
\end{equation}
where  $\mathbf{s} = [s_1,\cdots,s_K]^T\in \mathbb{R}^{K\times1}$ can be viewed as the extra interference caused by the channel estimation error.
It is very difficult to rigorously prove the robust beamforming structure with outage probabilistic constraints,
so a mathematical interpretation of  \eqref{eq:w} is provided in the Appendix at the end of this paper.
We denote robust beamforming matrix as $\hat \qW = [\hat \qw_1,\cdots,\hat \qw_K]^T$.
It is worth noting that \eqref{eq:w}, although not rigorously mathematically derived, is able to approximate the value of $\qs_k$ through a neural network description without the need for its exact expression.
This structure enables us to reconstruct the robust beamforming matrix $\hat \qW$ from the output $[\qp,\qq,\qs]$ of the neural network when only the estimated CSI is known, and the output dimension of the neural network is reduced from $2N K$\footnote{Since neural network frameworks such as PyTorch or TensorFlow have difficulty dealing with complex numbers, the complex channel must be divided into real and imaginary parts.} to $3K$.
\subsection{Bipartite GNN for Feature Learning}\label{Sec-GNN}
In our previous works \cite{dl_bf,icc}, power features $[\qp,\qq]$ can be learned by a fully connected neural network (FCNN) or CNN.
Considering that GNNs can achieve state of the art performance, and the potential scalability as well as generalizability due to their permutation equivariant and permutation invariant properties \cite{GNN,GNN2,GNN3,GNN3-1},
this paper employs a special bipartite GNN (BGNN) as in \cite{GNN} to realize the BMP inference and thus to learn the features in \eqref{eq:w}.
	Different with \cite{GNN}, the weights of the edges in the BGNN are estimated channel information $\tilde{\qH}$ rather than perfect channel information $\qH$.
	Besides, we need not only to learn the power feature $[\qp,\qq]$ with a BGNN, but also to learn the interference feature $\qs$ with another BGNN considering the effect of channel errors.
	The reason of using two BGNNs to learn the interference feature $\qs$ and  power features $[\qp,\qq]$ separately is due to the difference in their ranges of values.
	For narrative simplicity, in this subsection we use the notation $\qg_k$ to denote the features of user $k$ learned by the BGNN, which either stands for being $\qs_k\in \mathcal R$ or $[\qp_k,\qq_k]\in \mathcal R^2$.

\begin{figure*}[h]
	\centering
	\includegraphics[width=0.9\linewidth]{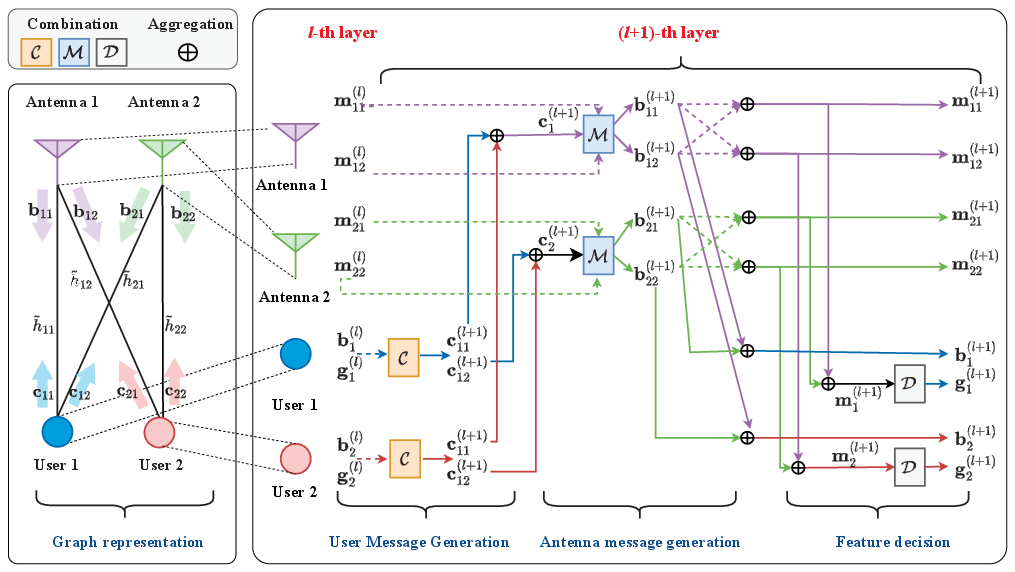}
	\caption{Graph representation  and deep learning based BMP  inference with $N=2, K=2$.}
	\label{fig: BMP}
\end{figure*} 

The MU-MISO system can be represented as  a weighted bipartite graph $\mathcal{G} = (\mathcal{N},\mathcal{K},\tilde\qH)$,
where the $\mathcal{N}$ and $\mathcal{K}$ respectively stand for disjoint vertex sets of antennas and users, and estimated channel information $\tilde \qH$ refers to the weight matrix of all edges,
as shown in Fig. \ref{fig: BMP}.
Each antenna-user pair $(n,k), \forall n\in \mathcal{N}, \forall k\in\mathcal{K}$ is connected through an edge with weight $\tilde h_{nk}$.
At this point, we realize the division of the basic network entity for the MU-MISO system by considering each antenna of the BS and the user as a vertex.
In this paper, we adopt a deep learning method based on vertices parameter sharing \cite{GNN,GNN4},
which facilitates the GNN to satisfy the desired permutation equivariance properties and is irrelevant to how the graph is constructed.
The basic idea is that each vertex can complete its own computation independently based on part of the estimated channel information, and then transfer the relevant statistics (named as messages) between the vertices to avoid the lack of overall channel information. By iterating the this process, the overall computation of the system can be formed, and finally the optimal beamforming of the MU-MISO system can be realized.

 The iteration process of vertex parameter  sharing  for beamforming feature calculation can be achieved through the BMP inference from the literature \cite{GNN}, which defines three types of vertex operators, i.e., user message generators $\mathcal{C}(\cdot)$, antenna message generators $\mathcal{M}(\cdot)$, and feature decision makers $\mathcal{D}(\cdot)$.
We use an $L$-layer BGNN to realize the BMP inference, where each layer of the BGNN implements an iteration in the BMP inference, as shown in Fig. \ref{fig: BMP}.
In the following, we will briefly introduce the deep learning based BMP inference,  see \cite{GNN} for more details.

Denote the BGNN as $\mathbf \Phi(\tilde{\qH};\mathbf \Theta)$ with the trainable parameter set $ \mathbf \Theta$.
In $l+1$-th layer of the BGNN, vertex operators $\mathcal{C}(\cdot)$, $\mathcal{M}(\cdot)$, and $\mathcal{D}(\cdot)$ are respectively implemented by three FCNNs $\mathbf \Phi_{\mathcal{C}_{l+1}}(\cdot; \mathbf \Theta_{\mathcal{C}_{l+1}})$, $\mathbf \Phi_{\mathcal{M}_{l+1}}(\cdot;\mathbf \Theta_{\mathcal{M}_{l+1}})$ and $\mathbf \Phi_{\mathcal{D}_{l+1}}(\cdot;\mathbf \Theta_{\mathcal{D}_{l+1}})$ as follows:

\begin{equation}
	\qc_{kn}^{(l+1)}= \mathcal{C}\bigg(\qg_k^{(l)}, \qb_k^{(l)},\tilde h_{ki}\bigg)=\mathbf \Phi_{\mathcal{C}_{l+1}}\bigg(\qg_k^{(l)}, \qb_k^{(l)},\tilde h_{kn};\mathbf \Theta_{\mathcal{C}_{l+1}}\bigg), \label{eq: ckn}
\end{equation}
\begin{eqnarray}
	\qb_{nk}^{(l+1)} &&= \mathcal{M}\bigg(\qm_{nk}^{(l)}, \qc_n^{(l+1)} ,\tilde h_{kn}\bigg)\notag\\
	&&=\mathbf \Phi_{\mathcal{M}_{l+1}}\bigg(\qm_{nk}^{(l)}, \qc_n^{(l+1)} ,\tilde h_{kn};\mathbf \Theta_{\mathcal{M}_{l+1}}\bigg)\label{eq: bnk},
\end{eqnarray}
\begin{equation}
	\qg_k^{(l+1)} = \mathcal{D}\bigg(\qm_k^{(l+1)}\bigg) = \mathbf \Phi_{\mathcal{D}_{l+1}}\bigg(\qm_k^{(l+1)}; \Theta_{\mathcal{D}_{l+1}}\bigg), \label{eq: g}
\end{equation}
where the superscript $(\cdot)^{(l)}$ represents the value of $l$-th layer.
In the $l+1$-th layer, each user vertex has a copy of sub neural networks $\mathbf \Phi_{\mathcal{C}_{l+1}}(\cdot; \mathbf \Theta_{\mathcal{C}_{l+1}})$ and  $\mathbf \Phi_{\mathcal{D}_{l+1}}(\cdot;\mathbf \Theta_{\mathcal{D}_{l+1}})$,
while each antenna vertex shares the same sub neural network $\mathbf \Phi_{\mathcal{M}_{l+1}}(\cdot;\mathbf \Theta_{\mathcal{M}_{l+1}})$.
Therefore, the trainable parameter set of the BGNN can be denoted as
\begin{equation}
 \mathbf \Theta = \big\{\mathbf \Theta_{\mathcal{C}_{l}},	\mathbf \Theta_{\mathcal{M}_{l}}, \mathbf \Theta_{\mathcal{D}_{l}},  l = 1,\cdots,L\big\} .
\end{equation}
$\qb_k^{(l+1)},\qm_{nk}^{(l+1)},\qc_n^{(l+1)},$ and $\qm_k^{(l+1)}$ are given below
\begin{eqnarray}
	&&\qb_k^{(l+1)} = \mathcal{P} (\{\qb_{nk}^{(l+1)},\forall n\in \mathcal{N}\}), \label{eq: bk}	\\
	&&\qm_{nk}^{(l+1)}=\{\qb_{nk}^{(l+1)},\mathcal P(\{\qb_{nj}^{(l+1)},\forall j\neq k\})\},\label{eq: mnk}\\
	&&\qc_n^{(l+1)} = \mathcal{P}(\{\qc_{kn}^{(l+1)},\forall n\in \mathcal{K}\}), \label{eq: cn} \\
	&&\qm_k^{(l+1)} = \mathcal{P}(\{\qm_{nk}^{(l+1)},\forall n\in \mathcal{N}\}) \label{eq: mk},
\end{eqnarray}
where $\mathcal P(\cdot)$ stands for the well-known sum pooling operator, which is widely used in existing works \cite{Pooling 2,Pooling3}.
Specifically, let $\mu_a$ be a variable of vertex $a\in \mathcal{A}$,
and we use a simplified pooling operator as $\mathcal{P}(\{\mu_a,\forall a\in \mathcal{A}\}) = \sum_{a\in \mathcal{A}} \mu_a$.

\begin{figure*}[h]
	\centering
	\includegraphics[width=0.7\linewidth]{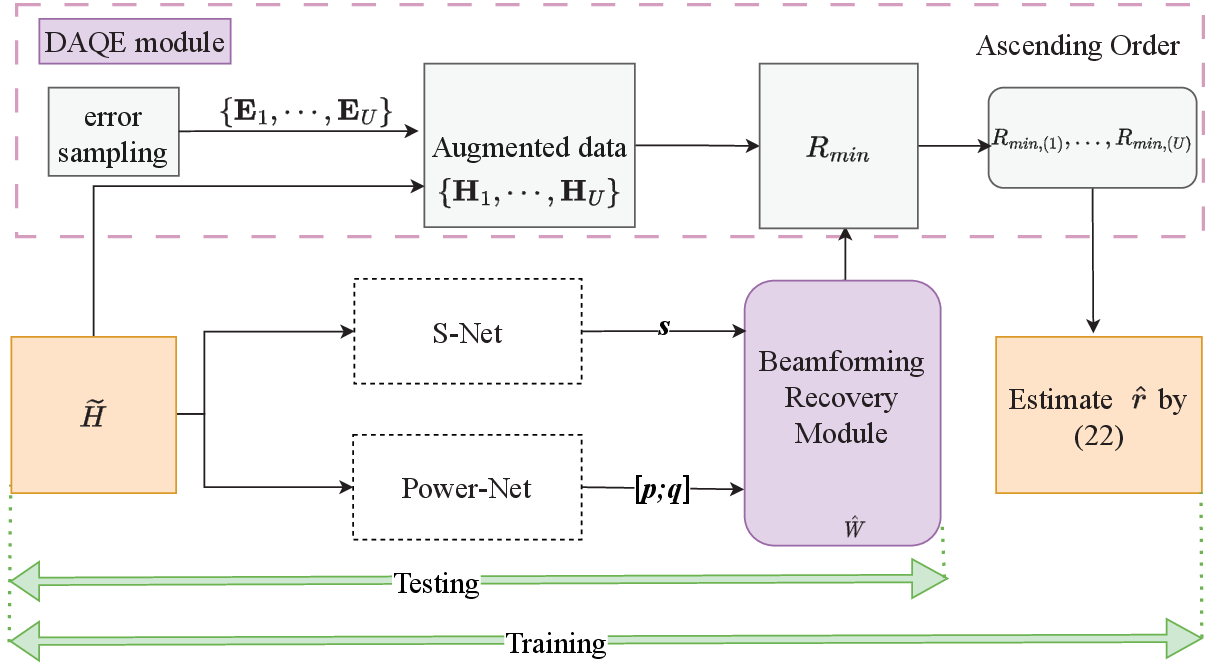}
	\caption{The proposed overall neural network framework for $\mathbf \Psi_1(\hat \qH; \mathbf \Theta)$.}
	\label{fig:DNN}
\end{figure*}

There are three main steps in $l+1$-th layer of the BGNN, which are detailed as follows:

(1) User message generation.
The user message $\qc_{kn}^{(l+1)} \in \mathbb{R}^M$  is generated at user vertex $k$ by \eqref{eq: ckn} and will be forwarded to the antenna $n$,
where the hyperparameter $M$ is the dimension of this message.
The neural network $\mathbf \Phi_{\mathcal{C}_{l+1}}(\cdot; \mathbf \Theta_{\mathcal{C}_{l+1}})$ realizes a combine operation of  the previous feature decision $\qg_k^{(l)}$, antenna message $\qb_k^{(l)}\in \mathbb{R}^{M}$ and $\tilde h_{kn}$.
Input and output dimensions of the neural network $\mathbf \Phi_{\mathcal{C}_{l+1}}(\cdot; \mathbf \Theta_{\mathcal{C}_{l+1}})$  are $M+2+dim(\qg_k^{(l)})$ and $M$, where $dim(\qg_k^{(l)}) = 2$ for power features $[\qp_k,\qq_k]$ and $dim(\qg_k^{(l)}) = 1$ for the interference feature $\qs_k$.
Therefore, the size of neural network  $\mathbf \Phi_{\mathcal{C}_{l+1}}(\cdot; \mathbf \Theta_{\mathcal{C}_{l+1}})$  is decoupled with the network size $N$ and $K$.

(2) Antenna message generation.
After receiving user messages $\{\qc_{kn},\forall k\in \mathcal{K}\}$,
the antenna vertex $n$ adopts the aggregation operation to obtain $\qc_n^{(l+1)} \in \mathbb{R}^M $ via \eqref{eq: cn}.
Then, it generates the antenna message $\qb_{nk}^{(l+1)}\in \mathbb{R}^{M}$ by \eqref{eq: bnk}, where the neural network $\mathbf \Phi_{\mathcal{M}_{l+1}}(\cdot;\mathbf \Theta_{\mathcal{M}_{l+1}})$ performs a combination operation on the previous antenna information $\qm_{nk}^{(l)}\in \mathbb{R}^{2M}$, $\qc_n^{(l+1)}$ and $\tilde h_{kn}$.
The input and output dimensions of $\mathbf \Phi_{\mathcal{M}_{l+1}}(\cdot;\mathbf \Theta_{\mathcal{M}_{l+1}})$ are $3M+2$ and $M$, respectively.
Subsequently, the antenna vertex $n$  updates $\qm_{nk}^{(l+1)}$ by \eqref{eq: mnk} and forwards $\qb_{nk}^{(l+1)}$ as well as $\qm_{nk}^{(l+1)}$  to the user vertex $k$.

(3) Feature decision.
The user $k$ aggregates all the received $\qm_{nk}^{(l+1)}, \forall n \in \mathcal{N}$ to obtain $\qm_k^{(l+1)} \in  \mathbb{R}^{2M}$  by \eqref{eq: mk}, and then generates the feature decision $\qg_k^{(l+1)} $ through $\mathbf \Phi_{\mathcal{D}_{l+1}}(\cdot;\mathbf \Theta_{\mathcal{D}_{l+1}})$ as in \eqref{eq: g},  without considering the network size $N$ and $K$.
Besides, it updates $\qb_k^{(l+1)}$ by the aggregation operation  according to \eqref{eq: bk}.

 Since each layer of the BGNN realizes one iteration of the BMP inference, through several times of iterations (e.g., $L$), user vertices are able to finally obtain the converged $\qg_k,\forall k\in\mathcal K$, which are used to recover the robust beamforming matrix~\eqref{eq:w}.

\subsection{Overall Neural Network framework}
After introducing the BGNN , we provide the overall neural network framework to address the outage constrained rate quantile maximization problem $\textbf{P1}$,
as shown in Fig. \ref{fig:DNN}.
Specifically, we provide a data  augmentation method to estimate the rate quantile and to gain robust beamforming solutions.
The input of the proposed network is the estimated CSI $\mathbf{\tilde H}$ with a dimension of $2N K$ by splitting the real part  $Re(\mathbf{\tilde H})$ and the imaginary part $Im(\mathbf{\tilde H})$,
and the outputs are the key features of robust beamforming $[\qs,\qp,\qq]$ with a dimension of $3K$ in total.
These features are used to recover beamforming solutions, and together with MC sampling to  estimate rate quantile $\hat r$ which satisfied the constraint \eqref{eq: OutageConstraint}.
Four main modules comprise the proposed network, which are detailed in the following:

\begin{itemize}
	\item \textbf{\textit{S-Net}}. This network is designed to learn the interference component caused by the channel estimation errors, i.e., $\qs\in \mathbb R^K$.
	It can be implemented by the aforementioned BGNN in Section \ref{Sec-GNN}  and
	is denoted as $\mathbf \Phi_{\mathcal{S}}(\tilde{\qH};\mathbf \Theta_{\mathcal{S}})$ with the parameter set $\mathbf \Theta_{\mathcal{S}} = \big\{\mathbf \Theta_{\mathcal{C}_{l}}^{\mathcal S},	\mathbf \Theta_{\mathcal{M}_{l}}^{\mathcal S}, \mathbf \Theta_{\mathcal{D}_{l}}^{\mathcal S} \forall l = 1,\cdots,L\big\}$.
	The hyperparameter $M$ in \textit{S-Net} is denoted $M_{\mathcal S}$.
	The details of the adopted BGNN will be given in Section \ref{Sec:Results}.
	
	\item  \textbf{\textit{Power-Net}}.
	Another BGNN is used to form the \textit{ Power-Net}, which aims to learn the power features of the beamforming (namely, $\qp \text{ and } \qq$). To distinguish it from the \textit{S-Net}, we denote \textit{Power-Net} as $\mathbf \Phi_{\mathcal{P}}(\tilde{\qH};\mathbf \Theta_{\mathcal{P}})$ with the parameter set $\mathbf \Theta_{\mathcal{P}} = \big\{\mathbf \Theta_{\mathcal{C}_{l}}^{\mathcal P},	\mathbf \Theta_{\mathcal{M}_{l}}^{\mathcal P}, \mathbf \Theta_{\mathcal{D}_{l}}^{\mathcal P} \forall l = 1,\cdots,L\big\}$.
	The hyperparameter $M$ in \textit{P-Net} is denoted $M_{\mathcal P}$.

	\item \textbf{\textit{Beamforming Recovery (BR) Module}}.
	The purpose of this module is to find the robust beamforming matrix $\hat{\mathbf{W}}$ based on \eqref{eq:w} from the downlink and virtual uplink power $\qp \text{ and } \qq$ obtained in \textit{Power-Net}, the interference feature $\qs$ obtained in \textit{S-Net}, and the estimated CSI $\mathbf{\tilde H}$.
	Note that this module does not contain parameters to be trained and optimized.

	\item \textbf{\textit{Data Augmentation based Quantile Estimation (DAQE) Module}}.
	This module is used to evaluate the rate quantile $\hat r$.
	When we calculate the output  $\hat{\mathbf{W}}$  in the BR module, the perfect channel information $\mathbf{H}$ is unknown, which fails to realize the robust design of beamforming.
	To this end, we need to calculate the outage performance to obtain the value of loss function and to evaluate the neural network's performance, but the challenge lies in that the outage constraint \eqref{eq: OutageConstraint} does not have a closed-form expression.
	A natural idea to solve this challenge is that we may adopt  MC sampling to estimate  the value of rate quantile,
	which has been widely used in conventional convex optimization methods as in \cite{Samplebased1,Samplebase2}.
	This inspired us to provide the so called data augmentation based quantile estimation method, as described in Algorithm \ref{alg:DAQE}.
	Specifically, we generate a potential channel error set $\{\qE_1,\cdots,\qE_U\}$ with $U$ elements through sampling from a certain distribution, e.g., circularly symmetric complex Gaussian distribution with zero mean and variance $\sigma _e^2$.
	This combined with the estimated CSI  $\mathbf{\tilde H}$ forms a set of potential actual channel $\{\qH_1,\cdots,\qH_U\}$ via \eqref{eq: realChannel}.
	Then we can calculate all the possible rate set $\{R_{k,u}, \forall k\in\mathcal{K}, u=1,\cdots,U\}$ given beamforming matrix  $\hat{\mathbf{W}}$  and  the potentially actual channel set  $\{\qH_1,\cdots,\qH_U\}$ through \eqref{eq: rate}.
	This is followed by a minimum operation among all users and the minimum rate set $\{R_{min,1},\cdots,R_{min, U}\}$ is obtained.
	After that, we sort them in ascending order and yield the set $\{R_{min,(1)},\cdots,R_{min, (U)}\}$ , where $R_{min, (u)}$ denotes the $u$-th smallest element in this set. 
	Thus, we can estimate the value of rate quantile $\hat r$ via a linear interpolation method.
	To be specific, after mapping $\rho \in (0,1)$ to the sample indices $[0,U]$, we can find  the location of quantile between two samples
	with indices $\lfloor \rho U \rfloor$ and $\lceil \rho U \rceil$ in the sorted order, and then we can obtain the rate quantile by
	\begin{equation}
		\hat r =  R_{min, (\lfloor U\rho \rfloor)} + \beta( R_{min, (\lceil U\rho \rceil)}- R_{min, (\lfloor U\rho \rfloor)}), \label{eq: quantile}
	\end{equation}
	where $\beta$ is the fractional part of the computed quantile index $\rho U$.
	In our previous studies, such methods combined with FCNNs have been applied to the beamforming problem with SINR probabilistic constraints \cite{icc}, and to the beamforming problem with SINR and harvested energy probabilistic constraints in SWIPT systems \cite{PIEEE}.
	
\end{itemize}

\begin{algorithm}[!t]
	\caption{Data Augmentation based Quantile Estimation}
	\label{alg:DAQE}
	\small
	\begin{algorithmic}[1]
		\STATE Obtain the estimated channel information $\tilde{\qH}$ and the learned beamforming matrix $\hat \qW$.
		\STATE Generate $U$ channel error samples $\{\qE_1,\cdots,\qE_U \}	\overset{iid}{\sim} \mathcal{CN}(0,\sigma _e^2)$.
		\STATE Obtain potential actual CSI set $\{\qH_1,\cdots,\qH_U\}$ by \eqref{eq: realChannel} and then calculate the corresponding rate set $\{R_{k,u}, \forall k\in\mathcal{K}, u=1,\cdots,U\}$ by \eqref{eq: rate} with learned $\hat \qW$.
		\STATE Find the minimum rate set $\{R_{min,1},\cdots,R_{min, U}\}$ among all users, and then sort them in  ascending order and yield the set $\{R_{min,(1)},\cdots,R_{min, (U)}\}$ , where $R_{min, (u)}$ denotes the $u$-th smallest element in this set.
		\STATE Estimate the rate quantile $\hat r$ by \eqref{eq: quantile}.
	\end{algorithmic}
\end{algorithm}

Now we could summarize the training and testing strategies to the proposed neural network.
The loss function is defined as follows:
\begin{equation}\label{eq:Loss}
	\text{Loss}= -\hat{r} = - \tau\big(\mathbf{\Psi}_1(\tilde{\qH};\mathbf{\Theta})\big),
\end{equation}
where $\mathbf{\Theta}=\{\mathbf\Theta_{\mathcal{S}},\mathbf\Theta_{\mathcal{P}}\}$ is the overall parameter set. The function $\tau(\cdot)$ defines a mapping relation from $\hat \qW$ to $\hat r$ and represents the quantile estimation process of the DAQE module.

At the $t$-th step of the training stage,
we  generate a mini-batch set $\mathcal H $  with the cardinality of $|\mathcal H|$ as  the input of the neural network,
which contains   the estimated CSI $\tilde{\qH}$.
The \textit{S-Net} and \textit{Power-Net} are computed in parallel to get the beamforming feature $[\qs,\qp,\qq]$
which together with the estimated CSI form the robust beamforming matrix $\hat{\mathbf{W}}$.
Then, the estimated rate quantile $\hat r$ can be obtained through the DAQE module,
and $-\hat r$ is used as the loss function defined in \eqref{eq:Loss} to train the overall neural network by the back-propagation algorithm.
The parameter set $\mathbf \Theta$ can be trained by stochastic gradient descent (SGD) algorithm \cite{SGD} or its variants, such as the Adam algorithm \cite{Adam}, which is updated as follows:
\begin{equation}
\mathbf \Theta^{(t)} = \mathbf \Theta^{(t-1)} + \frac{\alpha}{|\mathcal H |} \sum_{\tilde \qH \in \mathcal H} \tau \big(\nabla_{\mathbf{\Theta}} \mathbf \Psi_1\big(\tilde \qH;\mathbf \Theta^{(t-1)}\big)\big),
\end{equation}
where $\alpha$ represents the learning rate of the neural network.
Next, we will briefly discuss differentiability of rate quantile $\hat r$.
The rate function defined in \eqref{eq: rate} is differentiable with respect to the output of neural network $[\qs,\qp,\qq]$, which is also differentiable with respect to the parameter $\mathbf \Theta$.
The MC-sampling based the quantile value is  a convex combination of the $\lfloor U\rho \rfloor$-th and $\lceil U\rho \rceil$-th smallest values of sampled min-rate value,
which will not change the differentiability.
Therefore, the estimated quantile is differentiable according to the chain rule of differentiation.

When it comes to the testing stage, we can obtain the beamforming matrix via forward propagation in the well trained neural network.
The DAQE module will  not be involved at this stage, which means that we do not need to perform MC sampling to obtain this robust beamforming matrix, reducing the latency of the algorithm.
The computational complexity of neural network inference is analyzed as follows. Assume that the FCNN on each vertex has only one hidden layer with number of neurons $z$. The computational complexity of a BGNN is $O(LNKzM)$.
	The computational complexity of beamforming recovery by \eqref{eq:w} is dominated by the inverse operation, which is $O(KN^3)$.
	Therefore, the total inference complexity for problem \textbf{P1} is $O(LNKz(M_{\mathcal S} + M_{\mathcal P}) + KN^3)$.

\section{Power Minimization using Bisection Algorithm}\label{Sec:PowerMin}
Using deep learning method to deal with the power minimization problem with the outage-based constraint (i.e., \textbf{P2}) is more challenging,
because it  encounters MC sampling of many variables and frequent Lagrange multipliers updating, which increase the computational and latency overhead of training stage,
as  in our previous work \cite{icc}.
Another problem need to be addressed is that traditional deep learning methods are trained for a fixed transmit power, resulting in the lack of scalability of the neural network.
In practice, a fixed power is unrealistic, and should vary within a certain range in accordance with different
network topologies and power optimization schemes.
For instance, the heterogeneous network may includes macro-, micro- and pico-cells
 with different transmit power budgets \cite{HetNetCells}, and a BS will adaptively adjust its maximum transmit power according to its access user number and service requirements to reduce energy consumption.
To this end, we first extend the above-mentioned neural network framework to incorporate the feature of universal power.
Then, we propose a bisection-based approach to the power minimization problem.

\subsection{Neural Network with Universal Power}
\begin{figure}[h]
	\centering
	\includegraphics[width=\linewidth]{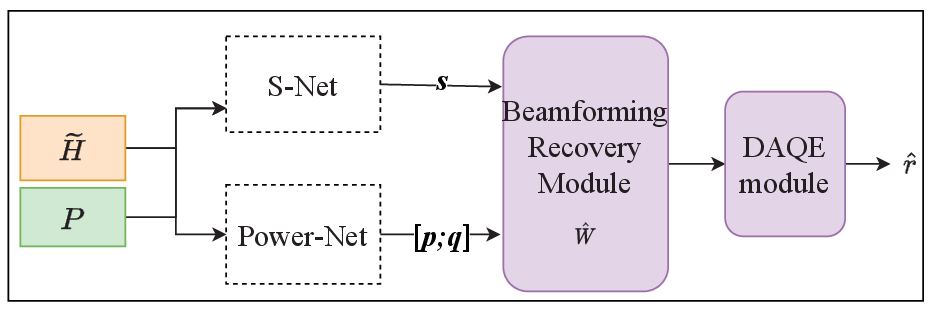}
	\caption{Modified framework for $\mathbf \Psi_2(\tilde \qH,P;\mathbf \Theta)$.}
	\label{fig: ModifiedNet}
\end{figure}
To achieve robust beamforming optimization with universal power budget,
an intuitive but inefficient idea is that we could train multiple DNNs for various transmission power level.
This practice obviously suffers from high computational complexity and storage, thus hindering the implementation of these DNNs.
To address this challenge, we take the transmit power as an additional input for the proposed neural network, which is similar as the work for beamforming design with perfect CSI \cite{Universal-power}.
The modified neural network can be represented as follows:

\begin{equation}
\{\qw_1,\cdots,\qw_K\} =\mathbf \Psi_2(\tilde \qH,P;\mathbf \Theta),
\end{equation}
where $\mathbf \Psi_2(\tilde \qH,P;\mathbf \Theta)$ has a similar network framework as $\mathbf \Psi_1(\tilde \qH;\mathbf \Theta)$, except that the transmit power budget is added as an input. The Loss function of $\mathbf \Psi_2(\tilde \qH,P;\mathbf \Theta)$ is $\text{Loss}= -\hat{r} = - \tau\big(\mathbf{\Psi}_2(\tilde{\qH};\mathbf{\Theta})\big)$.
The architecture of the modified neural network $\mathbf \Psi_2(\tilde \qH,P;\mathbf \Theta)$ is shown in Fig. \ref{fig: ModifiedNet}.

To be specific, inputs of the modified neural network include the estimated CSI and the transmit power, with the total dimension of $2NK+1$.
Transmit power can be viewed as a random variable and is uniformly sampled from a specific range of power variations,
namely, $P\sim \text{Uniform}(\underline{P},\bar{P})$, where $\underline{P} \text{ and } \bar{P}$ denote the potential minimum and maximum transmit power level, respectively.
Then, $[\qs, \qp,\qq]$ can be learned from the \textit{S-Net} and  \textit{Power-Net} with a total output dimension of $3K$ for the universal power.
Note that in order to satisfy the power constraint, the output of  \textit{Power-Net}  needs to be normalized
, i.e.,  $\sum_{k=1}^{K} p_k = \sum_{k=1}^{K} q_k  = 1$, which are then scaled to the total transmit power $P$ in the beamforming recovery module.
In this way, we can evaluate the maximum rate quantile $\hat{r}$ that stands for the maximum achievable rate while satisfying the outage constraint.

\subsection{Bisection Algorithm}

Since the sum rate of the MU-MISO system increases monotonically with the increasing transmission power of the BS,
we can use the bisection method to find the minimum transmit power from the potential power region.
This motivates us to solve the power minimization problem \textbf{P2} by the bisection method with the aid of the improved neural network proposed in the previous sub-section, as shown in Algorithm \ref{alg:A}.
Let us first set a range of possible transmit power $[\underline{P},\bar{P}]$ and a minimum rate target $r$, and use a trained universal power neural network to infer  the maximum rate quantile $\hat r$. We utilize bisection to find the minimum power until the iterative algorithm stopping condition (e.g., $|\hat r-r|\leq \varepsilon$) is satisfied, where $\varepsilon$ is an arbitrarily small positive number.
It is worth noting that in this process, the neural network does not need to be retrained due to its power universal property, so this method only increases the computational overhead of the bisection method, which is often considered to be an efficient  method.
\begin{algorithm}[!t]
	\caption{The bisection method based power minimization algorithm}
	\label{alg:A}
	\small
	\begin{algorithmic}[1]
		\STATE  Initial upper and lower limits of total transmit power $\bar P$ and $\underline{P}$, target minimum rate $r$, the outage probability requirement $\rho$, and stopping threshold $\varepsilon$.
		\WHILE{$|\hat r-r|\leq \varepsilon$}
		\STATE $P = \frac{\underline{P}+\bar P}{2}$;
		\STATE Obtain $\hat r$  and $\hat \qW$ by the modified network ($\mathbf \Psi_2$) inference with input $P$ .
		\IF{$\hat r < r$}
		\STATE $P= \underline{P}$
		\ELSE
		\STATE $P= \bar{P}$
		\ENDIF
		\ENDWHILE
		\RETURN  $\qW^* = \hat \qW$.
		
	\end{algorithmic}
\end{algorithm}

Let us now summarize the training and testing strategies of the modified neural network to solve power minimization problem \textbf{P2}.
During the training stage, a mini-batch set $\mathcal{H}$ with the cardinality of $|\mathcal H|$ containing numerous estimated CSI and transmit power is generated for learning the parameter set,
which can be updated as follows:
\begin{equation}
	\mathbf \Theta^{(t)} = \mathbf \Theta^{(t-1)} + \frac{\alpha}{|\mathcal H |} \sum_{\{\tilde \qH,P\} \in \mathcal H} \nabla_{\mathbf{\Theta}} \tau \big(\mathbf \Psi_2\big(\tilde \qH,P;\mathbf \Theta^{(t-1)}\big)\big).
\end{equation}
Once the network  $\mathbf \Psi_2$ is trained, we can use it combined with bisection algorithm to infer the the solution of \textbf{P2}.
Note that unlike solving \textbf{P1}, the bisection method requires the use of estimated rate quantile $\hat r$ when solving problem \textbf{P2},
where MC sampling is essential and will increase the computation complexity.
The computation complexity of rate quantile estimation and bisection algorithm are $O(U\log_2(U)KN)$ and $\log_2\big(\frac{\bar{P}-\underline{P}}{\varepsilon}\big)$, respectively.
Therefore, the total complexities to solve problem \textbf{P2} is $O\big(\log_2\big(\frac{\bar{P}-\underline{P}}{\varepsilon}\big)\big[LNKz(M_{\mathcal S} + M_{\mathcal P}) + KN^3 + U\log_2(U)KN\big]\big)$.
However, the execution time of this method is still much shorter than that of traditional optimization algorithms, and this will be demonstrated in  Section \ref{Sec:Results}.

\section{Numerical Results}\label{Sec:Results}
In this section, we provide numerical results to validate and evaluate  the performance of the proposed deep learning-based robust beamforming approach.
For the sake of comparison, we adopt the same parameter setting as \cite{power-only-solution}.
Unless otherwise specified, we consider a  MU-MISO system consisting of $N=4$ transmit antennas and $K=4$ users.
Rayleigh fading is used to model the estimated MISO channel $\tilde{\qH}$, where each single link (from an antenna to a user) follows a circularly symmetric complex Gaussian distribution with zero mean and unit variance, namely, $h_{nk}\sim \mathcal{CN}(0,1),~  \forall n \in \mathcal{N}, \forall k\in \mathcal{K}$.
The total bandwidth $B = 10$ MHz, the transmission power $P=30$ dBm, and the variance of channel estimation error $\sigma_e^2 = 0.075$.
Considering path loss and inter-cell interference, we set the noise power spectral density (PSD) to -75 dBm/Hz.
The statistical performance requirement induced by the channel uncertainties is  set as the largest outage probability $\rho = 0.05$.
\begin{table}[!t]
	\centering
	\caption{The Architecture and Parameters of FCNNs.}
	\label{Tab-Snet}
	\begin{tabular}{l c c}
		\hline \midrule
		\multicolumn{3}{c} {$\mathbf \Phi_{\mathcal{C}_l}^{\mathcal S}(\cdot; \mathbf \Theta_{\mathcal{C}_l}^{\mathcal S})$.}\\[0.15cm] \hline
		Layers & Input dimension & Output dimension \\ \hline
		Dense + ReLu& $M_{\mathcal S} + 3$ & 200 \\ \hline
		Dense + Tanh & 200 & $M_{\mathcal S}$ \\ \hline \midrule
		
			\multicolumn{3}{c} {$\mathbf \Phi_{\mathcal{C}_l}^{\mathcal P}(\cdot; \mathbf \Theta_{\mathcal{C}_l}^{\mathcal P})$.}\\ [0.15cm] \hline
		Layers & Input dimension & Output dimension \\ \hline
		Dense + ReLu& $M_{\mathcal P} + 4$ & 200 \\ \hline
		Dense + Tanh & 200 & $M_{\mathcal P}$\\  \hline \midrule
		
			\multicolumn{3}{c} {$\mathbf \Phi_{\mathcal{M}_l}^{\mathcal S}(\cdot; \mathbf \Theta_{\mathcal{S}})$ and $\mathbf \Phi_{\mathcal{M}_l}^{\mathcal P}(\cdot; \mathbf \Theta_{\mathcal{P}})$.}\\ [0.15cm]\hline
		Layers & Input dimension & Output dimension \\ \hline
		Dense + ReLu& $3M+ 2$  & 200 \\ \hline
		Dense + Tanh & 200 & $M$\\ \hline
		\multicolumn{3}{l} {Note: $M= M_{\mathcal S}$ for $\mathbf \Phi_{\mathcal{M}_l}^{\mathcal S}$, and $M= M_{\mathcal P}$ for $\mathbf \Phi_{\mathcal{M}_l}^{\mathcal P}$ }\\ \hline \midrule
		
		\multicolumn{3}{c} {$\mathbf \Phi_{\mathcal{D}_l}^{\mathcal S}(\cdot; \mathbf \Theta_{\mathcal{D}_l}^{\mathcal S})$.}\\ [0.15cm] \hline
		Layers & Input dimension & Output dimension \\ \hline
		Dense + ReLu& $2M_{\mathcal S}+1$ & 200 \\ \hline
		Dense  & 200 & 1 \\ \hline \midrule
		
		\multicolumn{3}{c} {$\mathbf \Phi_{\mathcal{D}_l}^{\mathcal P}(\cdot; \mathbf \Theta_{\mathcal{D}_l}^{\mathcal P})$.}\\ [0.15cm] \hline
		Layers & Input dimension & Output dimension \\ \hline
		Dense + ReLu& $2M_{\mathcal P}+1$ & 200 \\ \hline
		Dense + Softmax & 200 & 2 \\ \hline \hline
	\end{tabular}
\end{table}
All simulation results are generated by using a computer with an Intel i7-7700 CPU and an NVIDIA Titan Xp GPU.
Note that we also used the neural network design with universal power (same as the method in Section \ref{Sec:PowerMin}) for $\mathbf \Psi_1(\hat \qH; \mathbf \Theta)$ in our simulations to allow it to adapt to different transmit powers.
We use two BGNNs with $L=5$ layers to respectively implement  \textit{S-Net} and \textit{Power-Net}, where the FCNN  is adopted to construct the DNN of all vertices.
The architecture and parameters of neural networks are given in \textbf{TABLE \ref{Tab-Snet}}.
The hyperparameters $M_{\mathcal S} $ and $M_{\mathcal P} $ are set as 3 and 5, respectively.
The rectified linear unit (ReLU)  function is used to be the activation function at all hidden layers of FCNNs.
The activation function at the hidden layer of all DNNs is the rectified linear unit (ReLU) function.
The hyperbolic tangent (Tanh)  and Softmax functions are used to be  the activation functions at the output layer of FCNNs accordingly.
Since we take the transmit power as an additional input, the input dimensions of $\mathbf \Phi_{\mathcal{D}_l}^{\mathcal S}(\cdot; \mathbf \Theta_{\mathcal{D}_l}^{\mathcal S})$ and $\mathbf \Phi_{\mathcal{D}_l}^{\mathcal P}(\cdot; \mathbf \Theta_{\mathcal{D}_l}^{\mathcal P})$ are to $2M_{\mathcal S}+1$ and $2M_{\mathcal P}+1$, respectively.
The initialization of  messages defined in Section \ref{Sec-GNN} is realized by independently sampling from the Gaussian distribution with zero mean and unit variance.

The neural network $\mathbf \Psi_1(\tilde \qH;\mathbf \Theta)$ for problem \textbf{P1} is trained using over 150 epochs with early stopping according to the validation at the end of each epoch. The batch size is 100. $10^5$ training data and $2\times 10^3$ testing data are used.  To realize the data augmentation, $U=1000$ channel errors are sampled from the circularly symmetric complex Gaussian distribution,  which then form the 1000 true channels  based on \eqref{eq: realChannel}.
Then, we can obtain the rate quantile estimation $\hat r$ through the DAQE module.
The $P$ is uniformly sampled a special power range, i.e.,  $P\sim \text{Uniform}(\underline{P},\bar{P})$.
We set  $\underline{P} = 0$ dBm and $\bar{P} = 35$ dBm.
Adam algorithm \cite{Adam} with a learning rate $\alpha = 10^{-3}$ is used for the training of both neural networks.
The training of $\mathbf \Psi_2(\tilde \qH,P;\mathbf \Theta)$ for problem \textbf{P2} is similar with that of $\mathbf \Psi_1(\tilde \qH;\mathbf \Theta)$.

The following benchmark schemes are considered for comparison:

\begin{itemize}
  \item \textbf{The traditional optimization approach}:  specifically, we use BTI method as described in \cite{tsp-solution} to obtain the solution via CVX \cite{CVX}, which achieves the superior robust performance among existing methods.
  \item \textbf{The  traditional optimization by evaluation}: note that the obtained solution of \cite{tsp-solution} is conservative, and therefore the achievable rate performance evaluated by the MC method is usually better than the optimized solution. We use  `by Evaluation' to denote the actual achievable rate performance.
  \item \textbf{The DNN with power optimization only approach} \cite{power-only-solution}: it adopts RZF beamforming structure to reduce the output dimension of neural network.
  The rate quantile is also obtained by MC sampling.
  \item \textbf{The DNN method}: the neural network directly learns the $\hat \qW$  rather than the features of any beamforming structure, which  is a purely data-driven method and requires frequent updating the Lagrange multipliers.
  \item \textbf{The proposed solution with $s_k =0 $}:  to evaluate the impact of the interference features in \eqref{eq:w} on the communication performance, we set $s_k,\forall k\in \mathcal K$ to 0 in \textit{S-Net} of the proposed method.
\end{itemize}
\subsection{Results of Rate Quantile Maximization Problem \textbf{P1}}
\begin{figure}[!t]
\centering
 \includegraphics[width=2.8in]{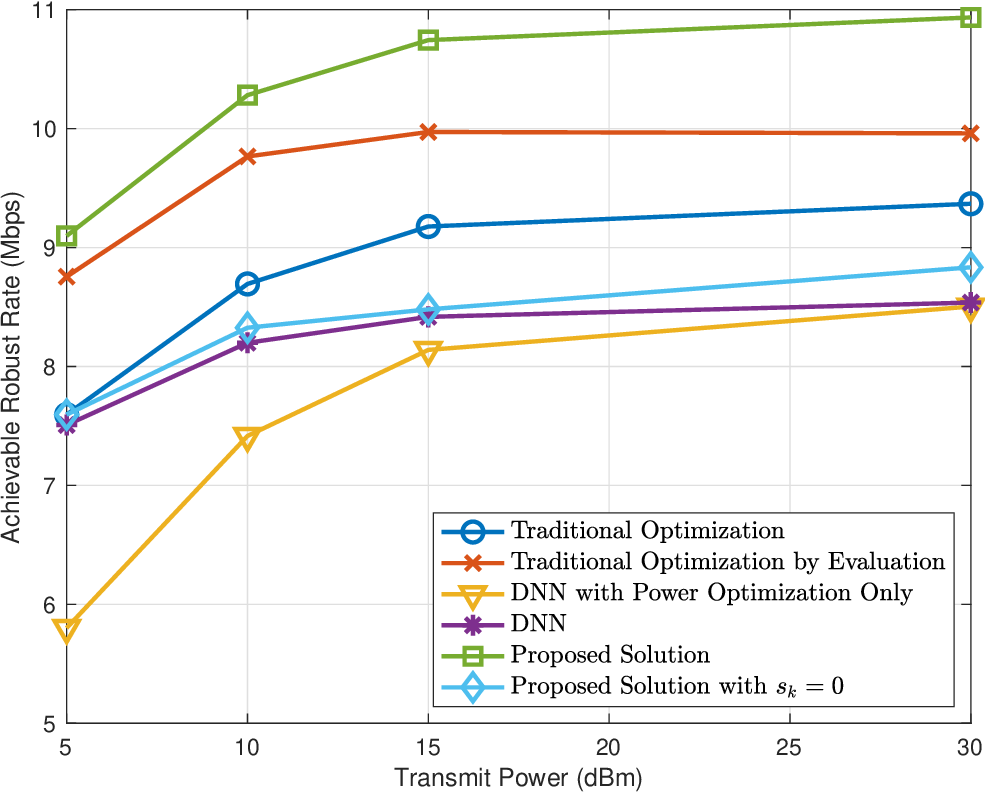}
  \caption{The achievable robust data rate against the transmit power.}
\label{fig:rate:vs:power}
\end{figure}
 
Fig. \ref{fig:rate:vs:power} shows that the achievable robust data rate increases with the increasing transmit power.  The proposed method achieves the highest robust data rate among all considered methods.
For instance, when the transmit power equals 30 dBm, the robust rate of the proposed method reaches about 10.9 Mbps, around 14\% higher than the traditional optimization method.
The reason for this result is that the traditional optimization methods use convex approximations such as convex restrictions and semidefinite programming, which loses the global optimality.
Interestingly, the traditional optimization algorithm actually computes a conservative solution which actually performs better.
We adopt the MC method to evaluate the traditional optimization method and found that it achieves about 0.7 Mbps higher data rate than the traditional optimization method, but still lower than the proposed method.
The proposed method with $s_k = 0$ obtains a lower robust rate performance, which implies that learning interference features (i.e., $s_k,\forall \mathcal K$) can indeed achieve performance improvement.
The  rate of the DNN method is much lower than above four methods because it only use data driven learning and thus is hard to improve the performance of the neural network.
The DNN with power optimization only method recovers the beamforming matrix via the learned power features and RZF beamforming model. Since the RZF beamforming only focuses on interference cancellation and ignores the impact of noise.
In contrast, other methods use  either the optimal structure or the rate-oriented optimization, and thus the DNN with power optimization only method obtains the lowest robust data rate.

\begin{figure}[!t]
\centering
 \includegraphics[width=2.8in]{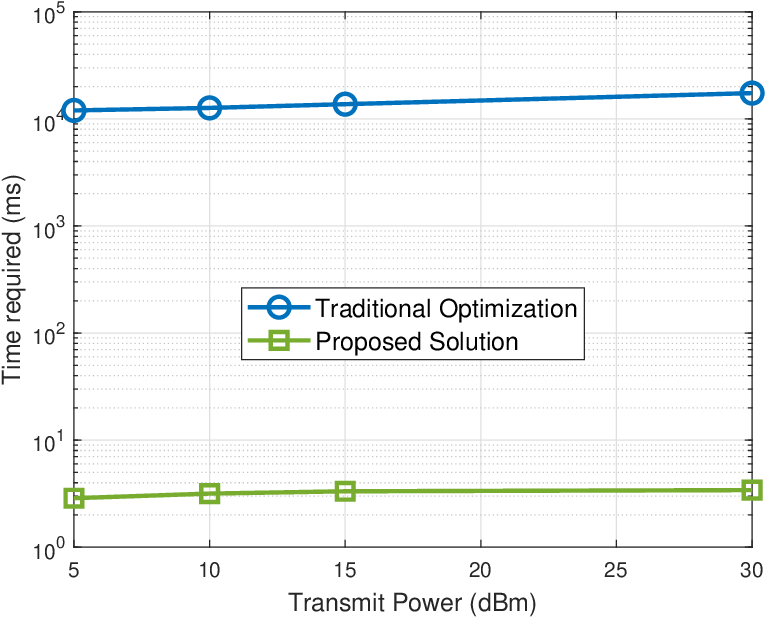}
  \caption{Comparison of execution time of different rate maximization algorithms.}
\label{fig:time:vs:power}
\end{figure}

Fig.  \ref{fig:time:vs:power} depicts the comparison between the traditional optimization method and our proposed method in terms of execution time under various transmit power.
Our proposed method has a fairly low execution latency of about 10 to 15 ms, which is about 1000 times lower than the traditional optimization method.
The reasons for this phenomenon lie in multiple folds.
Our proposed method only needs to use the learned network to infer the optimal solution.
In contrast, the traditional optimization method involves many iterations to obtain the solution, and loads of matrix operations (such as finding the relaxation matrices for convex restriction) in each iteration.
Once it obtains the rank-1 solution, the beamforming vectors for each user is obtained via matrix factorization due to the  limits of semidefinite programming.
If the solution suffers from high-rank, the Gaussian randomization procedure is required to obtain approximate beamforming vectors,
which further deteriorates the rate performance, the algorithm's execution time and even feasibility of the solution.
\textbf{Table \ref{tab:rank:SINR} }shows the probability of getting high-rank solutions for the rate quantile maximization problem,
and we found that this probability rises with increasing transmit power.
When $P=30$ dBm, obtaining a high-rank solution is a high probability event, i.e. 89.65\%.
This is in line with our previous observations in Fig.  \ref{fig:rate:vs:power} and Fig. \ref{fig:time:vs:power} about the deterioration of data rate performance and algorithm execution latency due to high-rank solutions for traditional optimization methods.
In these cases, we choose the beamforming directions as the principal  eigenvectors of the
high-rank solution, and then optimize the power only.

\begin{figure}[!t]
\centering
 \includegraphics[width=3.2 in]{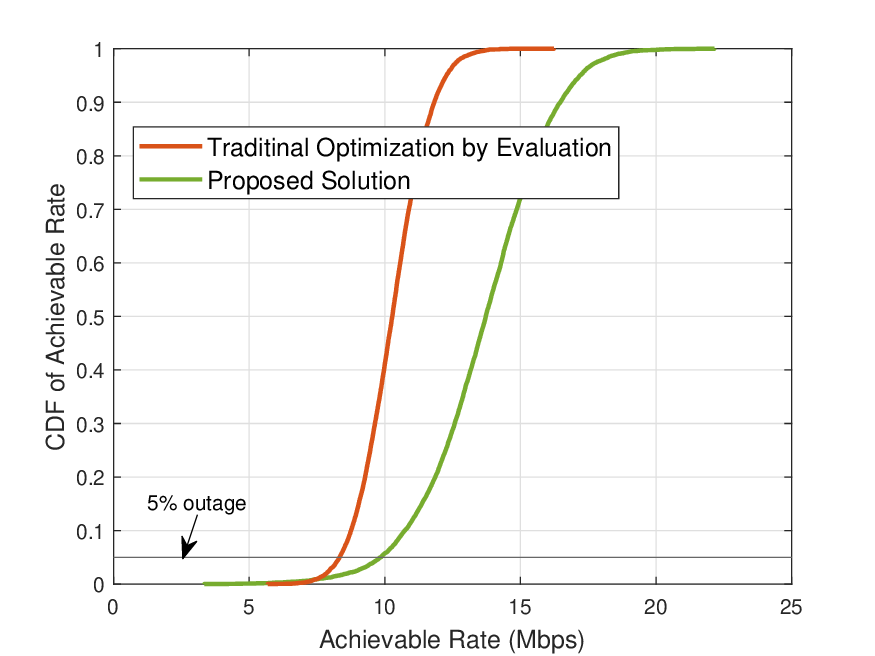}
  \caption{CDF of the achievable robust data rate for the rate maximization solution.}
\label{fig:cdf:rate:vs:power}
\end{figure}

Fig. \ref{fig:cdf:rate:vs:power} compares the cumulative distribution function (CDF) of achievable robust rate between the proposed solution and the traditional optimization approach by evaluation.
Since the traditional optimization approach may suffer from the high-rank solutions,
for comparative fairness, we choose a channel realization that
the traditional approach gives a rank-1 solution.
Then, to evaluate the robust rate, we generate 2000 channel error samples to estimate the robust rate at outage probability of 5\%.
The result shows that given the 5\% outage constraint, the achievable rate of the proposed solution reaches 10 Mbps,  about 2 Mbps higher than that of the traditional optimization approach by evaluation.
From another point of view, if a 10 Mbps achievable robust rate is desired,
the outage probability of traditional method  is as high as about 40\%,
where such solutions cannot be applied in a practical system.
In contrary, the proposed solution can satisfy the outage requirement 5\%.
It is worth emphasizing that the comparison environment we set up is very favorable to traditional optimization methods, which avoid the high-rank challenge, as well as using MC estimation instead of the original algorithmic results employed in the comparison of most works.
Despite this, our proposed method still achieves better performance.

\begin{table}[]
	\centering
	\caption{The probability of high-rank solution for the rate maximization problem.}
	\label{tab:rank:SINR}
	\begin{tabular}{|l|l|l|l|l| }
		\hline
		Power &  5 & 10  & 15 & 30   \\ \hline
		Probability&  0.2520 & 0.6375 &  0.8060  &  0.8965 \\ \hline
	\end{tabular}
\end{table}

\begin{figure}[!t]
	\centering
	\includegraphics[width=2.8in]{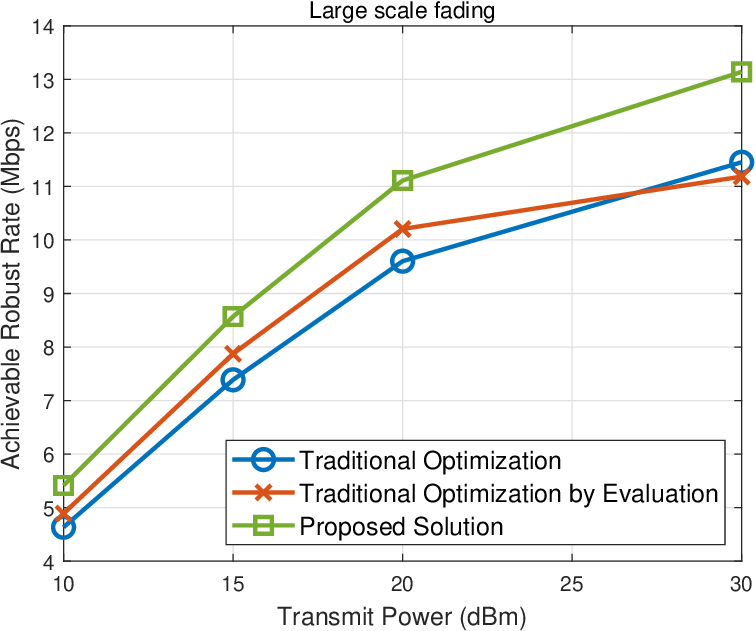}
	\caption{The achievable robust data rate against the transmit power with large scale fading.}
	\label{fig:rate:vs:power:large}
\end{figure}

In the above simulations, the path loss and inter-cell interference are combined to PSD (i.e., -75 dBm), where a similar approach is used in \cite{tsp-solution} and \cite{power-only-solution}. Fig. \ref{fig:rate:vs:power:large} shows the achievable robust data rate with the increasing transmit power under large-scale fading.
Specifically, the path loss is modeled as $128.1 + 37.6\log_{10}(d)$ where $d$ is the distance in kilometer between the BS and a user. Users are uniformly located in an area between 100 and 500 meters from the BS. The small-scale fading follows the Rayleigh distribution with zero mean and unit variance.  Simulation results again show that our proposed method is able to achieve better robust rate performance.

\begin{figure}[t]
	\centering
	\includegraphics[width= 2.8 in]{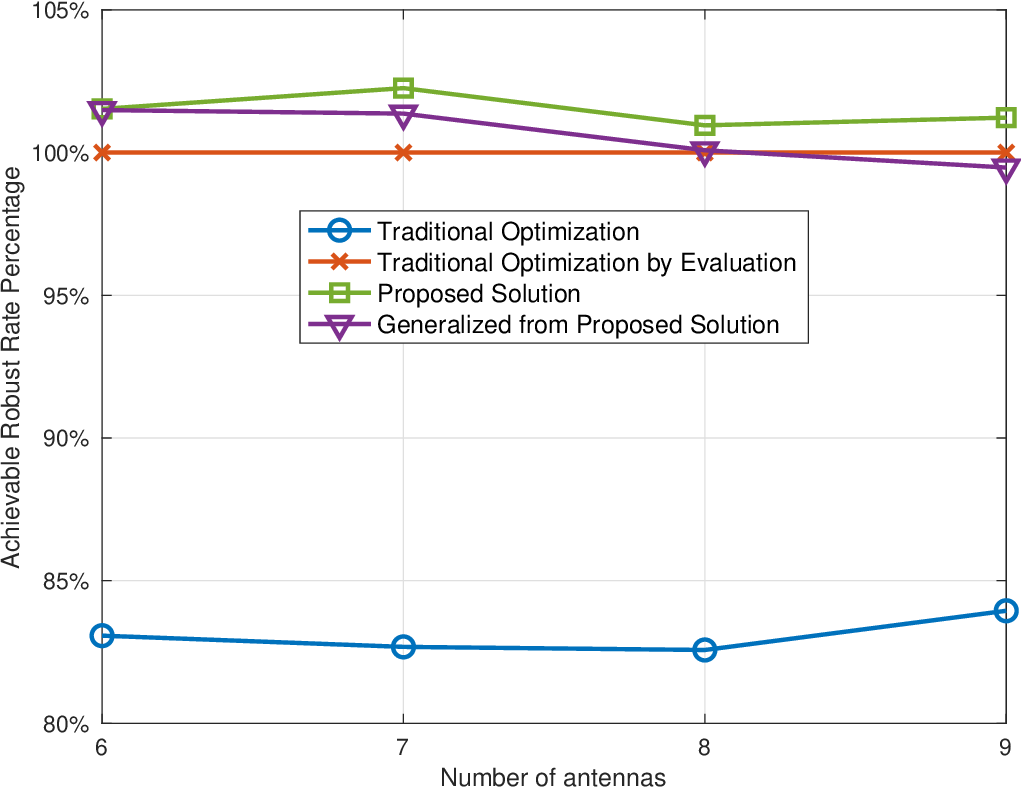}
	\caption {The scalability and generalizability of the proposed method.}
	\label{fig:rate:vs:Nt:percentage}
\end{figure}

Fig. \ref{fig:rate:vs:Nt:percentage} depicts the scalability and generalizability of the proposed method for a 4-user system, where the number of transmit antennas is varied from 6 to 9.
	$M_{\mathcal  S}=3$ and $M_{\mathcal  P}=8$ are chosen as hyperparameters for the neural network.
	The robust rates of all methods are normalized by the rate of the traditional optimization by evaluation method for the same system parameters,
	and the corresponding ratios are expressed as percentages.
	Our proposed method consistently outperforms traditional optimization (and by evaluation) methods in terms of robust rate performance under different sizes of communication systems, demonstrating excellent scalability.
	In order to describe the generalizability of the proposed method, we first train the neural network for the system size of  $N=6,K=4$, and use the trained network for a larger number of transmit antennas at the BS, where systems with $N\in [7,9]$ were not trained by the neural network.
The curve marked with ``generalized from proposed solution'' describes the generalizability performance of the above simulation.
	It can be seen that the robust beamforming rate of the generalized from proposed solution outperforms traditional optimization (and by evaluation) methods at $N=7\text{ and }8$. At $N=9$, it achieves only slightly worse performance than the traditional optimization by evaluation method, but still outperforms  the traditional optimization method.
\subsection{Results of Power Minimization Problem \textbf{P2}}
\begin{figure}[!t]
 	\centering
 	\includegraphics[width=2.8in]{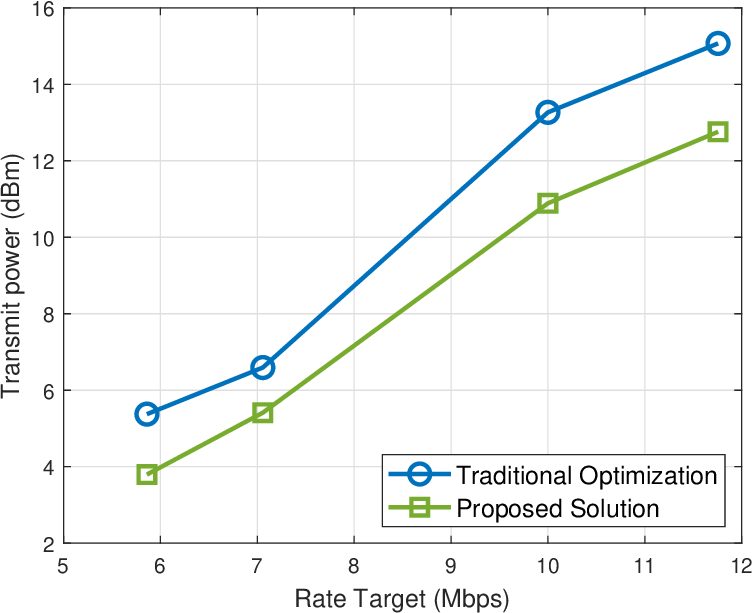}
 	\caption{The minimum transmit power against the data rate requirements.}
 	\label{fig:power:vs:rate}
 \end{figure}

Fig. \ref{fig:power:vs:rate} shows the transmit power versus required robust rate requirements.
With the increasing rate requirements, larger transmit power is needed.
The proposed method saves more power compared with the traditional optimization method.
For instance, when rate target is equal to 10 Mbps, the power consumption of the proposed method is around 11 dBm,  2.2 dBm lower than the traditional optimization method.
 In many cases, there is no feasible solution for given rate and outage constraints, so the power shown is only averaged over those feasible solutions.

\begin{figure}[!t]
	\centering
	\includegraphics[width=2.8in]{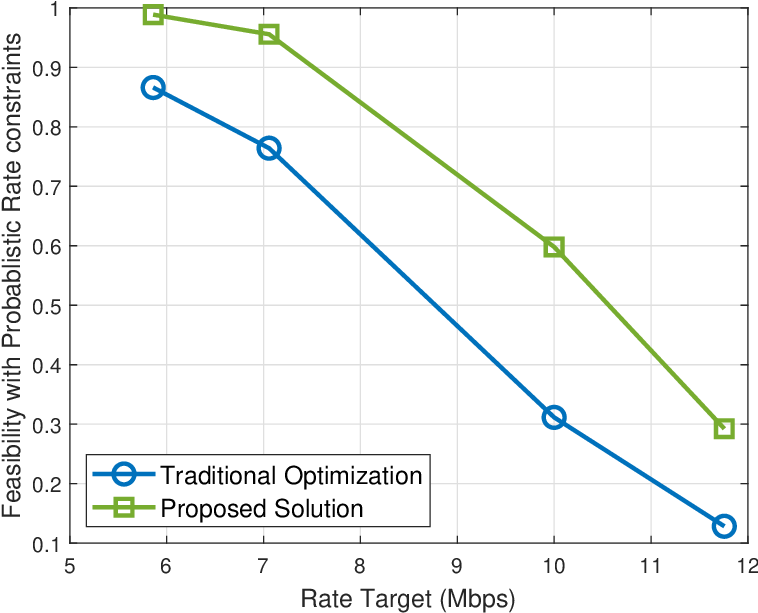}
	\caption{Feasibility probability of the solutions versus rate target.}
	\label{fig:feasibility_vs_rate}
\end{figure}

 Fig. \ref{fig:feasibility_vs_rate} depicts the probability of obtaining feasible solutions which satisfy the outage constraint in \textbf{P2}.
  	The feasibility of both methods decreases with the increasing target rate,	because higher  rate targets will cause more outage.
  	The robust beamforming solutions of our proposed approach have much higher feasibility probabilities than the traditional optimization approach.
  	For instance, when the rate target is equal to 10 Mbps, the proposed approach can get robust solutions with about 60\% feasibility, 30\% higher than the traditional optimization method.
  	This is because the solution obtained by the traditional optimization method using convex restriction and semidefinite relaxation is  conservative and thus this method restricts the solution space and reduces the chance of obtaining feasible solutions.
  	In addition, when high-rank solutions are obtained by semidefinite relaxation, it is in general difficult to find a feasible beamforming solution that satisfies the outage probability constraints.
  	In contrast, our method utilizes neural networks to optimize the robust beamforming, which can explore a much larger solution space and produce less conservative solutions, thus making it easier to ensuring feasibility.

\begin{figure}[!t]
\centering
 \includegraphics[width=2.8in]{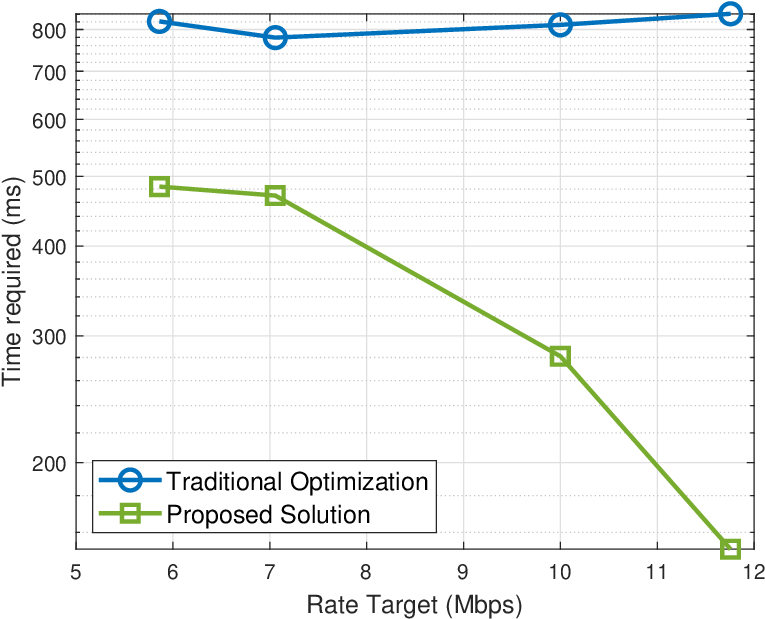}
  \caption{Comparison of execution time of different power minimization algorithms.}
\label{fig:time:vs:rate}
\end{figure}

Fig. \ref{fig:time:vs:rate} shows the comparison of execution time between the proposed method and the traditional optimization method for the power minimization problem under different rate targets.
The proposed solution spends much more time on the power minimization problems than the rate maximization problem because it uses bisection search and in each iteration, and it requires MC method to evaluate the rate quantile value. However, it still spends much less time than the traditional optimization method.
For instance, when the rate target is equal to 10 Mbps, the execution time of the proposed method is about 500 ms less than that of the traditional optimization method.
Besides, the execution time of the proposed method decreases with the increasing target rates, because the bisection method can identify the infeasible solution early to avoid unnecessary iterations.
Table \ref{tab:rank:power} depicts the probability of high-rank solutions of the traditional optimization method for the power minimization problems.

\begin{figure}[!t]
	\centering
	\includegraphics[width=3.2in]{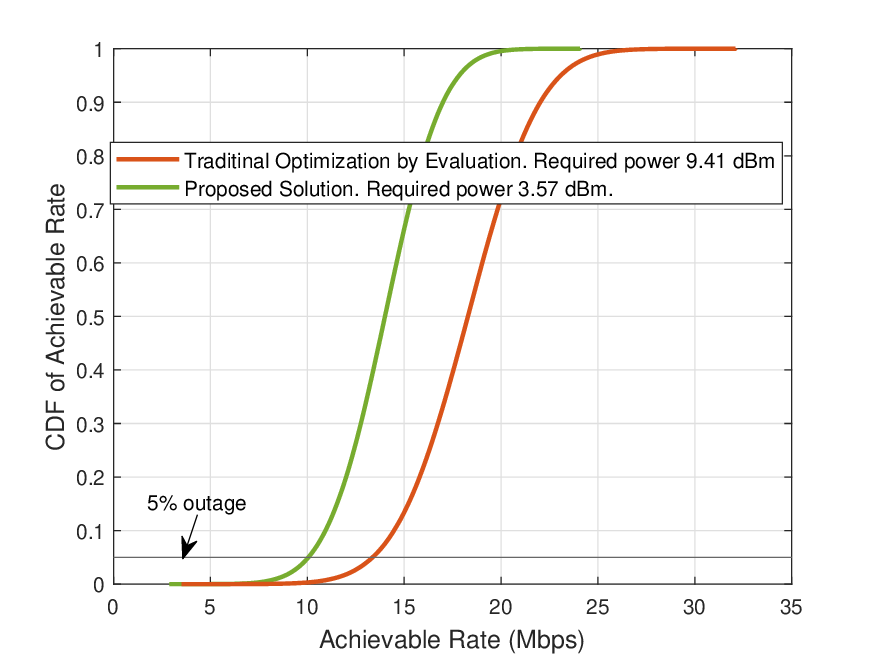}
	\caption{CDF of the achievable robust data rate for the power minimization solution.}
	\label{fig:cdf:rate:vs:power:min}
\end{figure}

In Fig. \ref{fig:cdf:rate:vs:power:min}, again we choose a channel for which the traditional optimization gives a rank-1 solution and compare the CDF and the power consumption. The rate requirement is 10 Mbps. The results are interesting: first, verifies that the traditional is conservative which means it achieves higher than 10 Mbps at outage of 5\%  while our proposed solution can achieve exactly 10 Mbps, avoiding waste of resources; second, our solution also uses nearly 6 dBm less power, which confirms its efficiency in using limited resources to achieve better and more robust solutions.

\begin{table}[]
	\centering
	\caption{The probability of high-rank solution for the power minimization problem.}
	\label{tab:rank:power}
	\begin{tabular}{|l|l|l|l|l| }
		\hline
		Rate (Mbps) &  5.86 & 7.06 & 10 & 11.76 \\ \hline
		Probability & 0.1149  & 0.1577 &0.1573   &  0.1479  \\ \hline
	\end{tabular}
\end{table}

\section{Conclusions}\label{Sec:Conclusion}
This paper proposed a data and model-driven learning approach to address robust beamforming optimization for the MU-MISO system, considering Gaussian channel uncertainties. The paper investigated two optimization problems with the outage probability constraint: the rate quantile maximization problem and the power minimum problem. For the rate quantile maximization problem, we proposed a model-driven learning approach combined with data augmentation to handle the challenging outage constraint. The modified robust beamforming structure is employed to reduce the neural network's output dimension from $2NK$ to $3K$. For the power minimum problem, we introduced a bisection-based learning method by incorporating the transmit power budget into the input of the proposed neural network, achieving the feature of universal transmit power. Numerical results demonstrated that the proposed approach outperforms state-of-the-art counterparts in terms of rate improvement, power reduction, and execution time.

\appendix[Mathematical Interpretation of \eqref{eq:w}]
The overall neural network framework illustrated in Fig. 2 of the manuscript solves the robust beamforming design based on outage constraints in the following two steps.

\begin{itemize}
	\item Neural networks (including \textit{S-Net} and \textit{Power-Net}) together with beamforming recovery model obtain optimal beamforming for generalized problems, where the outage probability constraint and  the minimum rate are not involved here.  Denote $\Gamma_k$ as the SINR at user $k$, and the generalized problems corresponding to \textbf{P1} and \textbf{P2} are respectively given as follows:
	\begin{eqnarray}
		\textbf{E1:}   \max_{\mathbf{w}_k,\forall \mathcal{K}} f(\Gamma_1,\cdots,\Gamma_K)\ \ \ \ \mbox{s.t.}\ \ \     \sum_{k=1}^{K} \Vert \mathbf{w}_k \Vert^2 \leq P,
	\end{eqnarray}
	\begin{eqnarray}
		\textbf{E2:}  \min_{\mathbf{w}_k,\forall \mathcal{K}} \sum_{k=1}^{K} \Vert \mathbf{w}_k \Vert^2 \ \ \ \ \mbox{s.t.} \ \ \  R_k \geq r,
	\end{eqnarray}
	where $f(\cdot)$ stands for an arbitrary utility function that is strictly increasing in each user's SINR as defined in \cite{bf-structure}.
	\item The DAQE module estimates the quantile of the minimum rate among users $\hat r$ which satisfies the outage constraint. By using  $-\hat r$ as the loss function, the neural network gradually learns to output beamforming that can satisfy the outage probability constraints during training until convergence.
\end{itemize}

We then explore expressions for the optimal solution structure for \textbf{E1} and \textbf{E2} taking into account the channel error. According to \cite{bf-structure0, bf-structure},  the optimal beamforming solution for \textbf{E2} is also the optimal solution for \textbf{E1}, so we use \textbf{E2} to deduce the robust optimal beamforming structure. The constraints of \textbf{E2} can be recast as:
\begin{eqnarray}
	&&\Gamma_k  = \frac{  |\mathbf{h}_k^\dag \mathbf{w}_k|^2}{\sum_{j\neq k} |\mathbf{h}_{k}^\dag\mathbf{w}_j|^2+\sigma^2_k} \geq 2^{r/B}-1 \triangleq \gamma\\
	\Leftrightarrow && \frac{1}{\gamma \sigma_k^2}|\mathbf{h}_k^\dag \mathbf{w}_k|^2 \geq \sum_{j\neq k} \frac{1}{ \sigma_k^2}|\mathbf{h}_{k}^\dag\mathbf{w}_j|^2+1.
\end{eqnarray}
Obviously, the strong duality and Karush-Kuhn-Tucker (KKT) conditions are held for problem \textbf{E2} \cite{kkt-strongdual}, so we consider the Lagrangian function as
\begin{eqnarray}
	&&\mathcal{L}(\qw_1,\cdots,\qw_K, q_1,\cdots,q_k)   \notag \\
	= && \sum_{k=1}^{K} \Vert \qw_k\Vert^2 +\sum_{k=1}^{K} q_k \bigg(  \sum_{j\neq k} \frac{1}{ \sigma_k^2}|\mathbf{h}_{k}^\dag\mathbf{w}_j|^2  +1\notag \\
	&& -  \frac{1}{\gamma \sigma_k^2}|\mathbf{h}_k^\dag \mathbf{w}_k|^2\bigg),
\end{eqnarray}
where $q_k, \forall \mathcal{K}$ is the Lagrange multiplier.
Based on KKT conditions, the optimal solution will satisfy $\frac{\partial \mathcal{L}}{\partial \qw_k} = 0$. Then we have
\begin{eqnarray}
	&& \qw_k + \sum_{j\neq k} \frac{q_j}{\sigma_j^2}\qh_j\qh_j^\dag \qw_k - \frac{q_k}{\gamma \sigma_k^2} \qh_k\qh_k^\dag \qw_k = \mathbf{0} \notag \\
	\overset{(a)}{\Leftrightarrow} &&  \qw_k^\dag + \sum_{j\neq k} \frac{q_j}{\sigma_j^2}\qw_k^\dag\qh_j^\dag\qh_j  - \frac{q_k}{\gamma \sigma_k^2} \qw_k^\dag\qh_k^\dag \qh_k  = \mathbf{0}^T \label{eq:transpose1}
\end{eqnarray}
where (a)  holds because the conjugate transpose is taken simultaneously for both sides of \eqref{eq:transpose1}.
Consider $\qh_j = \tilde \qh_j + \qe_j$, we have
\begin{eqnarray}
	\qh_j^\dag\qh_j  = \tilde \qh_j^\dag \tilde \qh_j +  \underbrace{ 2 Re \big(\tilde \qh_j^\dag \qe_j\big)    + \qe_j^\dag \qe_j}_{\text{denote as } \beta_j \in \mathcal {R} }. \label{eq:hh}
\end{eqnarray}
Substituting \eqref{eq:hh} into \eqref{eq:transpose1}, we obtain
\begin{eqnarray}
	&& \qw_k^\dag  + \sum_{j\neq k} \frac{q_j}{\sigma_j^2}\qw_k^\dag ( \tilde \qh_j^\dag \tilde \qh_j + \beta_j) - \frac{q_k}{\gamma \sigma_k^2} \qw_k^\dag( \tilde \qh_k^\dag \tilde \qh_k + \beta_k) \notag\\
	&&  = \mathbf{0}^T  \label{eq:8}\\
	\overset{(a)}{\Leftrightarrow} && \qw_k^\dag \bigg(\qI + \sum_{j=1}^{K} \frac{q_j}{\sigma_j^2}   \tilde \qh_j^\dag \tilde \qh_j  +\notag \\
	&& \underbrace{\bigg (\sum_{j=1}^{K} \frac{q_j}{\sigma_j^2}   \beta_j -\frac{q_k}{\sigma_k^2} \bigg(1+\frac{1}{\gamma}\bigg)\beta_k \bigg)}_{\text{denote as } s_k \in R} \qI \bigg) \notag \\
	 && = \frac{q_k}{\sigma_k^2}\bigg(1+\frac{1}{\gamma}\bigg)\qw_k^\dag \tilde \qh_k^\dag \tilde \qh_k  \label{eq:9}\\
	\overset{(b)}{\Leftrightarrow}&& \qw_k = \bigg( \big (1+s_k\big )\qI + \sum_{j=1}^{K} \frac{q_j}{\sigma_j^2}   \tilde \qh_j \tilde \qh_j^\dag  \bigg)^{-1} \tilde \qh_k \notag \\
	&& \times  \underbrace{\frac{q_k}{\sigma_k^2}\bigg(1+\frac{1}{\gamma}\bigg)  \tilde \qh_k^\dag  \qw_k}_{ = \text {scalar}},
\end{eqnarray}
where $(a)$ is held by adding term $\frac{q_k}{\sigma^2_k}\qw_k^\dag \tilde \qh_k^\dag \tilde \qh_k$ to both side of \eqref{eq:8}, and (b) is  established by taking   conjugate transpose of \eqref{eq:9} and left multiplying the inverse matrix of $\big (1+s_k\big )\qI + \sum_{j=1}^{K} \frac{q_j}{\sigma_j^2}   \tilde \qh_j \tilde \qh_j^\dag$.
Note that $\frac{q_k}{\sigma_k^2}\bigg(1+\frac{1}{\gamma}\bigg)  \tilde \qh_k^\dag  \qw_k$ is a scalar,  so the optimal $\qw_k$ must be parallel to $\bigg( \big (1+s_k\big )\qI + \sum_{j=1}^{K} \frac{q_j}{\sigma_j^2}   \tilde \qh_j \tilde \qh_j^\dag  \bigg)^{-1} \tilde \qh_k$.
In view of this, the optimal robust beamforming  vectors have the following structures:
\begin{eqnarray}
	\tilde\qw_k = \sqrt{p_k} \frac{\bigg((1+s_k)\mathbf{I} + \sum_{j=1}^K \frac{q_j}{\sigma_j^2} \tilde\qh_{j}\tilde\qh_{j}^\dag  \bigg)^{-1} \tilde \qh_k}{\bigg \Vert\bigg((1+s_k)\mathbf{I} + \sum_{j=1}^K \frac{q_j}{\sigma_j^2} \tilde\qh_{j}\tilde\qh_{j}^\dag  \bigg)^{-1} \tilde \qh_k \bigg \Vert},
\end{eqnarray}
where $p_k$ is the beamforming power for user $k$.


\begin{thebibliography}{99}

\bibitem{BF0}
G. Zheng, K. -K. Wong and T. -S. Ng, ``Throughput Maximization in Linear Multiuser MIMO-OFDM Downlink Systems,'' \emph{IEEE Trans. Veh. Technol.}, vol. 57, no. 3, pp. 1993-1998, May 2008.

\bibitem{SPM1}
D. Gesbert, M. Kountrious, C.-B. Chae, R. W. Heath, Jr., and T. Salzer, ``Shifting the MIMO Paradigm," \emph{IEEE Signal Processing Mag.}, vol. 24, no. 5, pp. 36-46, Sep. 2007

\bibitem{BF1}
Y. Zhang and C. Tepedelenlioglu, ``Transmit Beamforming with Power Adaptation in Downlink Multi-User Systems,'' \emph{IEEE Trans. Wireless Commun.}, vol. 9, no. 8, pp. 2424-2429, Aug. 2010.
\bibitem{BF2}
Q. Shi, M. Razaviyayn, M. Hong and Z. -Q. Luo, ``SINR Constrained Beamforming for a MIMO Multi-User Downlink System: Algorithms and Convergence Analysis,'' \emph{IEEE Trans. Signal Process}, vol. 64, no. 11, pp. 2920-2933, Jun. 2016.

\bibitem{WorstCase}
S. -J. Kim, A. Magnani, A. Mutapcic, S. P. Boyd and Z. -Q. Luo, ``Robust Beamforming via Worst-Case SINR Maximization,'' \emph{IEEE Trans. Signal Process}, vol. 56, no. 4, pp. 1539-1547, Apr. 2008.
\bibitem{WorstCase2}
Z. L. Yu, W. Ser, M. H. Er, Z. Gu and Y. Li, ``Robust Adaptive Beamformers Based on Worst-Case Optimization and Constraints on Magnitude Response,'' \emph{IEEE Trans. Signal Process}, vol. 57, no. 7, pp. 2615-2628, Jul. 2009.
\bibitem{WorstCase3}
C. Li, C. He, L. Jiang and F. Liu, ``Robust Beamforming Design for Max–Min SINR in MIMO Interference Channels,'' \emph{ IEEE Commun. Lett.}, vol. 20, no. 4, pp. 724-727, Apr. 2016.
\bibitem{AverageBobust}
B. Zhang, Z. He, K. Niu and L. Zhang, ``Robust Linear Beamforming for MIMO Relay Broadcast Channel With Limited Feedback,'' \emph{ IEEE Signal Process Lett.}, vol. 17, no. 2, pp. 209-212, Feb. 2010.
\bibitem{AverageBobust2}
M. Medra, Y. Huang and T. N. Davidson, ``Offset-Based Beamforming: A New Approach to Robust Downlink Transmission,'' \emph{IEEE Trans. Signal Process}, vol. 67, no. 1, pp. 70-82, 1 Jan. 2019.
\bibitem{outage1}
C. Lin, C. -J. Lu and W. -H. Chen, ``Outage-Constrained Coordinated Beamforming With Opportunistic Interference Cancellation,'' \emph{IEEE Trans. Signal Process}, vol. 62, no. 16, pp. 4311-4326, Aug. 2014.
\bibitem{outage2}
W. -C. Li, T. -H. Chang, C. Lin and C. -Y. Chi, ``Coordinated Beamforming for Multiuser MISO Interference Channel Under Rate Outage Constraints,'' \emph{IEEE Trans. Signal Process}, vol. 61, no. 5, pp. 1087-1103, Mar. 2013.



\bibitem{tsp-solution}
K. -Y. Wang, A. M. -C. So, T. -H. Chang, W. -K. Ma and C. -Y. Chi, ``Outage Constrained Robust Transmit Optimization for Multiuser MISO Downlinks: Tractable Approximations by Conic Optimization,'' \emph{IEEE Trans. Signal Process}, vol. 62, no. 21, pp. 5690-5705, Nov. 2014.

\bibitem{SDP1}
Z. -Q. Luo, W. -K. Ma, A. M. -C. So, Y. Ye and S. Zhang, ``Semidefinite Relaxation of Quadratic Optimization Problems,'' \emph{IEEE Signal Process Mag.}, vol. 27, no. 3, pp. 20-34, May 2010.

\bibitem{SDP2}
S. H. Low, ``Convex Relaxation of Optimal Power Flow—Part I: Formulations and Equivalence,'' \emph{IEEE Trans. Control Netw. Syst.}, vol. 1, no. 1, pp. 15-27, Mar. 2014.

\bibitem{irs}
Y. Sun, K. An, J. Luo, Y. Zhu, G. Zheng and S. Chatzinotas, ``Outage Constrained Robust Beamforming Optimization for Multiuser IRS-Assisted Anti-Jamming Communications With Incomplete Information,'' \emph{IEEE Internet Things J.}, vol. 9, no. 15, pp. 13298-13314, Aug. 2022.

\bibitem{swipt}
B. Su, Q. Ni and W. Yu, ``Robust Transmit Beamforming for SWIPT-Enabled Cooperative NOMA With Channel Uncertainties,'' \emph{IEEE Trans. Commun.}, vol. 67, no. 6, pp. 4381-4392, Jun. 2019.

\bibitem{HetNet}
K. -Y. Wang, N. Jacklin, Z. Ding and C. -Y. Chi, ``Robust {MISO Transmit} Optimization under Outage-Based QoS Constraints in Two-Tier Heterogeneous Networks,'' \emph{IEEE Trans. Wireless Commun.}, vol. 12, no. 4, pp. 1883-1897, Apr. 2013.

\bibitem{Satellite}
Y. Yan, K. An, B. Zhang, W. -P. Zhu, G. Ding and D. Guo, ``Outage-Constrained Robust Multigroup Multicast Beamforming for Satellite-Based Internet of Things Coexisting With Terrestrial Networks,''  \emph{IEEE Internet Things J.}, vol. 8, no. 10, pp. 8159-8172, May 2021.

\bibitem{Rada-communication}
A. Bazzi and M. Chafii, ``On Outage-Based Beamforming Design for Dual-Functional Radar-Communication 6G Systems,'' \emph{IEEE Trans. Wireless Commun.}, vol. 22, no. 8, pp. 5598-5612, Aug. 2023.

\bibitem{FedLearning}
Y. Zou, Z. Wang, X. Chen, H. Zhou and Y. Zhou, ``Knowledge-Guided Learning for Transceiver Design in Over-the-Air Federated Learning,"  \emph{IEEE Trans. Wireless Commun.}, vol. 22, no. 1, pp. 270-285, Jan. 2023.

\bibitem{Offloading}
S. Bi, L. Huang, H. Wang and Y. -J. A. Zhang, "Lyapunov-Guided Deep Reinforcement Learning for Stable Online Computation Offloading in Mobile-Edge Computing Networks," \emph{IEEE Trans. Wireless Commun.}, vol. 20, no. 11, pp. 7519-7537, Nov. 2021.

\bibitem{dl_bf}
W. Xia, G. Zheng, Y. Zhu, J. Zhang, J. Wang and A. P. Petropulu, ``A Deep Learning Framework for Optimization of MISO Downlink Beamforming,'' \emph{IEEE Trans. Commun.}, vol. 68, no. 3, pp. 1866-1880, Mar. 2020.


\bibitem{bf-structure}
E.~Bj\"{o}rnson, M.~Bengtsson, and B.~Ottersten, ``Optimal multiuser transmit  beamforming: {A} difficult problem with a simple solution structure,'' \emph{IEEE Signal Process Mag.}, vol. 31, no. 4, pp. 142-148, Jul. 2014.
\bibitem{Model_Driven}
J. Zhang, M. You, G. Zheng, I. Krikidis and L. Zhao, ``Model-Driven Learning for Generic MIMO Downlink Beamforming With Uplink Channel Information,'' \emph{IEEE Trans. Wireless Commun.}, vol. 21, no. 4, pp. 2368-2382, Apr. 2022.

\bibitem{Universal-power}
J. Kim, H. Lee, S. -E. Hong and S. -H. Park, ``Deep Learning Methods for Universal MISO Beamforming,"  \emph{IEEE Wireless Commun. Lett.}, vol. 9, no. 11, pp. 1894-1898, Nov. 2020.

\bibitem{GNN}
J. Kim, H. Lee, S. -E. Hong and S. -H. Park, ``A Bipartite Graph Neural Network Approach for Scalable Beamforming Optimization,'' \emph{IEEE Trans. Wireless Commun.}, vol. 22, no. 1, pp. 333-347, Jan. 2023.

\bibitem{ModelNew1}
Q. Hu, Y. Cai, Q. Shi, K. Xu, G. Yu and Z. Ding, ``Iterative Algorithm Induced Deep-Unfolding Neural Networks: Precoding Design for Multiuser MIMO Systems," \emph{IEEE Trans. Wireless Commun.}, vol. 20, no. 2, pp. 1394-1410, Feb. 2021.
\bibitem{WMMSE}
Q. Shi, M. Razaviyayn, Z. Luo, and C. He, ``An iteratively weighted MMSE approach to distributed sum-utility maximization for a MIMO interfering broadcast channel," IEEE Trans. Signal Process., vol. 59, no. 9, Sept. 2011.

\bibitem{ModelNew2}
W. Jin, J. Zhang, C. -K. Wen and S. Jin, ``Model-Driven Deep Learning for Hybrid Precoding in Millimeter Wave MU-MIMO System,"  \emph{IEEE Trans. Commun.}, vol. 71, no. 10, pp. 5862-5876, Oct. 2023.

\bibitem{ModelNew3}
J. Guo and C. Yang, ``A Model-based GNN for Learning Precoding,"  \emph{IEEE Trans. Wireless Commun.}, doi: 10.1109/TWC.2023.3336911.

\bibitem{MU-MIMO}
M. Zhang, J. Gao and C. Zhong, ``A Deep Learning-Based Framework for Low Complexity Multiuser MIMO Precoding Design,'' \emph{IEEE Trans. Wireless Commun.}, vol. 21, no. 12, pp. 11193-11206, Dec. 2022.

\bibitem{StatisticRobust}
J. Kim, H. Lee and S. -H. Park, ``Learning Robust Beamforming for MISO Downlink Systems,'' \emph{IEEE Commun. Lett.}, vol. 25, no. 6, pp. 1916-1920, Jun. 2021.

\bibitem{Multicast-unicast}
Z. Zhang, M. Tao and Y. -F. Liu, ``Learning to Beamform in Joint Multicast and Unicast Transmission With Imperfect CSI,'' \emph{IEEE Trans. Commun.}, vol. 71, no. 5, pp. 2711-2723, May 2023.

\bibitem{icc}
M. You, G. Zheng and H. Sun, ``A Data Augmentation based DNN Approach for Outage-Constrained Robust Beamforming,'' \emph{IEEE Int. Conf. Commun.}, pp. 1-5, 2021.

\bibitem{power-only-solution}
W. Cui and W. Yu, ``Uncertainty Injection: A Deep Learning Method for Robust Optimization,'' \emph{IEEE Trans. Wireless Commun.}, vol. 22, no. 11, pp. 7201-7213, Nov. 2023.


\bibitem{RZF}
C. B. Peel, B. M. Hochwald, and A. L. Swindlehurst, ``A Vector-perturbation Technique for Near-capacity Multiantenna Multiuser Communication-part {I}: Channel Inversion and Regularization,'' \emph{IEEE Trans. Commun.}, vol. 53, no. 1, pp. 195–202, Jan. 2005.

\bibitem{CBF}
C.-B. Chae, D. Mazzarese, N. Jindal, and R. W. Heath, Jr., ``Coordinated Beamforming with Limited Feedback in the MIMO Broadcast Channel," \emph{IEEE J. Sel. Areas Commun.}, vol. 26, no. 8, pp. 1505-1515, Oct. 2008.

\bibitem{UniversalApproximation}
K. Hornik, M. Stinchcombe, and H. White, ``Multilayer feedforward networks are universal approximators,'' \emph{Neural Netw.}, vol. 2, no. 5, pp. 359–366, Jul. 1989.

\bibitem{MPGNN}
Y. Shen, Y. Shi, J. Zhang, and K. B. Letaief, ``Graph neural networks for scalable radio resource management: Architecture design and theoretical analysis,'' \emph{IEEE J. Sel. Areas Commun.}, vol. 39, no. 1, pp. 101–115, Jan. 2021.

\bibitem{GNN2}
K. Xu, W. Hu, J. Leskovec, and S. Jegelka, ``How Powerful are Graph Neural Networks?'' \emph{Proc. Int. Conf. Learn. Representation}, May 2019.

\bibitem{GNN3}
Z. Wu, S. Pan, F. Chen, G. Long, C. Zhang and P. S. Yu, ``A Comprehensive Survey on Graph Neural Networks,'' \emph{IEEE Trans. Neural Netw. Learn. Syst.}, vol. 32, no. 1, pp. 4-24, Jan. 2021.

\bibitem{GNN3-1}
Y. Shen, Y. Shi, J. Zhang, and K. B. Letaief, ``Graph Neural Networks for Scalable Radio Resource Management: Architecture Design and Theoretical Analysis,'' \emph{IEEE J. Sel. Areas Commun.}, vol. 39, no. 1, pp. 101–115, Jan. 2021.

\bibitem{GNN4}
J. Guo and C. Yang, ``Learning Power Allocation for Multi-Cell-Multi-User Systems With Heterogeneous Graph Neural Networks,'' \emph{IEEE Trans. Wireless Commun.}, vol. 21, no. 2, pp. 884-897, Feb. 2022.

\bibitem{Pooling 2}
H. Lee, S. H. Lee, and T. Q. S. Quek, ``Learning Autonomy in Management of Wireless Random Networks,'' \emph{IEEE Trans. Wireless Commun.}, vol. 20, no. 12, pp. 8039–8053, Dec. 2021.
\bibitem{Pooling3}
A. Chowdhury, G. Verma, C. Rao, A. Swami, and S. Segarra, ``Unfolding WMMSE Using Graph Neural Networks for Efficient Power Allocation,'' \emph{IEEE Trans. Wireless Commun.}, vol. 20, no. 9, pp. 6004–6017, Sep. 2021.



\bibitem{Samplebased1}
B. Dai, Y.-F. Liu, and W. Yu, ``Optimized Base-Station Cache Allocation for Cloud Radio Access Network With Multicast Backhaul,'' \emph{IEEE J. Sel. Areas Commun.}, vol. 36, no. 8, pp. 1737–1750, Aug. 2018.

\bibitem{Samplebase2}
Y. Shi, J. Zhang, and K. B. Letaief, ``Optimal Stochastic Coordinated Beamforming for Wireless Cooperative Networks With CSI Uncertainty,'' \emph{IEEE Trans. Signal Process.}, vol. 63, no. 4, pp. 960–973, Feb. 2015.
\bibitem{PIEEE}
C. Psomas, M. You, K. Liang, G. Zheng and I. Krikidis, ``Design and Analysis of SWIPT With Safety Constraints,"  \emph{Proceedings of the IEEE}, vol. 110, no. 1, pp. 107-126, Jan. 2022.
\bibitem{SGD}
I. Sutskever, J. Martens, G. Dahl, and G. Hinton, ``On the importance of initialization and momentum in deep learning,''  \emph{Int. Conf. Mach. Learn.}, pp. 1139-1147, May 2013.

\bibitem{Adam}
D. Kingma and J. Ba, ``Adam: A method for stochastic optimization,'' \emph{Proc. Int. Conf. Learn. Represent. (ICLR)}, pp. 1–15, 2015.

\bibitem{HetNetCells}
A. Damnjanovic \emph{et al}., ``A survey on 3GPP heterogeneous networks,'' \emph{IEEE Wireless Commun.}, vol. 18, no. 3, pp. 10-21, Jun. 2011.
\bibitem{CVX}
Michael Grant and Stephen Boyd, ``CVX: Matlab software for disciplined convex programming, version 2.0 beta.'' http://cvxr.com/cvx, Sep. 2013.

\bibitem{bf-structure0}
Emil Bj\"{o}rnson and Eduard Jorswieck, ``Optimal resource allocation in coordinated multi-cell systems,'' {\em Foundations and Trends in Communications and Information Theory}, vol. 9, no. 2-3, pp. 113-381, 2013.

\bibitem{kkt-strongdual}
W. Yu and T. Lan, ``Transmitter optimization for the multi-antenna downlink with per-antenna power constraints,” IEEE Trans. Signal Processing, vol. 55, no. 6, pp. 2646-2660, 2007.


\end{thebibliography}
\end{document}